\documentclass[preprint,3p,12pt,sort&compress]{elsarticle}

\usepackage{mathrsfs}
\usepackage{amsmath}
\usepackage{stmaryrd}
\usepackage{bbding}
\usepackage{dcolumn}
\usepackage{graphicx}
\usepackage{amsfonts}
\usepackage{amssymb}
\usepackage{psfrag}
\usepackage{wrapfig}
\usepackage{subfigure}
\usepackage{makeidx}
\usepackage{bm}
\usepackage{epsf}
\usepackage{epsfig}
\usepackage{setspace}
\usepackage{graphicx}
\usepackage{epstopdf}
\usepackage{psfrag}
\usepackage{subfigure}
\usepackage{color}
\usepackage{enumerate}
\usepackage{verbatim} 
\usepackage{booktabs}
\usepackage[utf8]{inputenc}
\usepackage{hyperref}
\usepackage{algorithm}
\usepackage{algpseudocode}
\usepackage{xltabular}

\begin{document}

\title{Solving continuum and rarefied flows using differentiable programming}

\author[IMECH,UCAS]{Tianbai Xiao}
\ead{txiao@imech.ac.cn}

\address[IMECH]{State Key Laboratory of High Temperature Gas Dynamics and Centre for Interdisciplinary Research in Fluids, Institute of Mechanics, Chinese Academy of Sciences, Beijing, China}
\address[UCAS]{School of Engineering Science, University of Chinese Academy of Sciences, Beijing, China}

\begin{abstract}
    Accurate and efficient prediction of multi-scale flows remains a formidable challenge.
    Constructing theoretical models and numerical methods often involves the design and optimization of parameters.
    While gradient descent methods have been mainly manifested to shine in the wave of deep learning, composable automatic differentiation can advance scientific computing where the application of classical adjoint methods alone is infeasible or cumbersome.
    Differentiable programming provides a novel paradigm that unifies data structures and control flows and facilitates gradient-based optimization of parameters in a computer program.
    This paper addresses the notion and implementation of the first solution algorithm for multi-scale flow physics across continuum and rarefied regimes based on differentiable programming.
    The fully differentiable simulator provides a unified framework for the convergence of computational fluid dynamics and machine learning, i.e., scientific machine learning.
    Specifically, parameterized mechanical-neural flow models and numerical methods can be constructed for forward physical processes, while the parameters can be trained on the fly with the help of the gradients that are taken through the backward passes of the whole simulation program, a.k.a., end-to-end optimization.
    As a result, versatile data-driven modeling and simulation can be achieved for physics discovery, surrogate modeling, and simulation acceleration.
    The fundamentals and implementation of the solution algorithm are demonstrated in detail.
    Numerical experiments, including forward and inverse problems for hydrodynamic and kinetic equations, are presented to demonstrate the performance of the numerical method.
    The open-source codes to reproduce the numerical results are available under the MIT license\footnote{\url{https://github.com/vavrines/KitAD.jl}}. 
\end{abstract}


\maketitle

\begin{table}
    \centering
    \caption{Nomenclature.}
    \begin{tabular*}{16cm}{lll}
        \hline
        $\partial P$ & differentiable programming \\
        AD & automatic differentiation \\
        $t$ & time variable \\
        $\mathbf x$ & space variables $(x,y,z)$ \\
        $\mathbf v$ & particle velocity variables $(u,v,w)$ \\
        $f$ & particle distribution function \\
        $\mathcal Q$, $\mathcal R$ & collision and relaxation operators in the kinetic equation \\
        $\mathcal M$ & Maxwellian distribution function \\
        $\nu$, $\tau$ & relaxation frequency and time $(\tau=1/\nu)$ \\
        $m$ & molecular mass \\
        $k$ & Boltzmann constant \\
        $\rho$, $\mathbf V$, $T$, $E$ & macroscopic primitive variables \\
        $\psi$ & collision invariants \\
        $\mathbf W$ & macroscopic conservative variables \\
        $\mathbf P$, $\mathbf T$, $\mathbf q$ & stress tensor, deviator tensor and heat flux \\
        $\bar f$, $\bar{\mathbf W}$ & cell-averaged distribution function and conservative variables \\
        $\mathbf F^f$, $\mathbf F^W$ & numerical fluxes for distribution function and conservative variables \\
        $\mathbf u$ & generic representation of flow variables \\
        $\mathcal F$ & operator of fluxes and sources in semi-discrete equations \\
        $\mathcal C$ & operator of initial and boundary conditions in semi-discrete equations \\
        $\mathbf p$ & control parameters \\
        $C$ & cost function \\
        $g$ & function integrated in the cost function \\
        $\lambda$ & adjoint variable \\
        $\mathbf d$ & data points in the training set \\
        $\digamma$ & generic representation of numerical operation \\
        $\partial\digamma$ & Jacobian and associated VJP and JVP of $\digamma$ \\
        $\mathbf s$ & states in a sequence of operations \\
        $\mathbf t$, $\mathbf r$ & intermediate variables for derivation in forward- and reverse-mode ADs \\
        $\mathrm{NN}_{\boldsymbol\theta}$ & neural network with trainable parameters $\boldsymbol\theta$ \\
        $\mathcal L$ & function layer in a neural network \\
        $\boldsymbol\omega$, $\mathbf b$, $\phi$ & weights, biases and activation function in a neural network \\
        $\boldsymbol\alpha$ & trainable parameters in the mechanical model \\
        $\mathbf u^\mathrm{ref}$ & referenced flow solution \\
        $\epsilon$ & regularization parameter \\
        $\mu$ & dynamic viscosity coefficient \\
        $\chi$ & proportion of upwind contribution to numerical fluxes \\
        $\hat f$ & reconstructed distribution function at cell face \\
        $\sigma$ & sigmoid function \\
        $h$ & reduced distribution function \\
        \hline
    \end{tabular*}
    \label{table:nomenclature}
\end{table}
\newpage















\section{Introduction}

Gaseous flows are endowed with a multi-scale structure.
Sufficient separation of scales facilitates the development of theories of fluid dynamics at different scales.
At molecular mean free path, the Boltzmann equation can be employed to describe the flight and collision effects of individual particles, while the Navier-Stokes equations depict the collective behavior of the many-particle system upon the fluid element models \cite{tsien1946superaerodynamics}.
Intrigued by the well-known Hilbert’s 6th problem \cite{hilbert1902mathematical}, continuous efforts have been made to bridge the gaps between the models from different scales, e.g., the Hilbert expansions from a theoretical point of view \cite{sone2002kinetic} and asymptotic-preserving numerical methods \cite{jin2022asymptotic}. 
These approaches build a cross-scale path to represent the upscaling effects with reasonable asymptotics.
However, it remains a formidable challenge to recover a continuous spectrum of flow physics and, in particular, to provide a succinct and accurate description in the transition regime.

Modeling and simulation of flows is a task of intertwined forward and inverse problems.
Building reliable theoretical models and numerical methods requires a proper determination of design variables.
At the molecular mean free path, i.e., the mesoscopic scale, phenomenological parameters in the collision kernel of the Boltzmann equation need to be routinely calibrated by experiments to preserve correct transport coefficients \cite{cercignani1988boltzmann}.
At the macroscopic level, constitutive functions are required as the closure of the Navier-Stokes equations and extended hydrodynamic models \cite{rosenau1989extending}.
From a numerical perspective, a numerical scheme's success depends on the optimization of the parametric solution algorithm, e.g., the nonlinear weights in the reconstruction stencil, the ratio of central and upwind contributions in the numerical flux function, and the coefficients in the Butcher tableau of the Runge-Kutta integrator.
Such optimization is more challenging for multi-scale flows since models and algorithms that are optimal at one scale are not necessarily suitable for another.

The burgeoning discipline of machine learning, especially deep learning, widens the possibility of studying complex flows under extreme conditions that seemed beset with difficulties in the past.
Deep neural networks as large parametric models enable versatile modeling and simulation of flow physics, including efficient solution of high-dimensional differential equations \cite{raissi2019physics,han2018solving}, operator learning for mappings of functions and distributions \cite{lu2021learning,li2020fourier}, and data-driven discovery of non-equilibrium physics \cite{floryan2022data,pan2018data}.
To improve the prediction of flow physics for such models, a typical workflow is to construct an objective function to be optimized with respect to the trainable parameters (known as loss function in deep learning) following the supervised or unsupervised learning approach.
The gradient information is usually obtained by backpropagating the loss through a chain of matrix operations, and the optimization problem is subsequently solved using stochastic gradient descent and its variants incorporating momentum and adaptive learning rates \cite{goodfellow2016deep}.

The optimization of parametric flow models and numerical methods can be performed based on two paradigms.
The first idea is to build prior flow datasets through high-fidelity experiments or fine-scale simulations, followed by supervised learning.
This approach is called offline training, and it can be applied to any parametric functions and operators on discrete spatio-temporal sensors.
However, this type of training usually has no direct perception of the time-evolving processes and spatio-temporal coupling embedded in the fluid dynamic equations.
As a result, the training data may not be utilized efficiently, which requires greater access to costly, high-confidence data \cite{menghani2023efficient}.
On the other hand, the solution process of computational fluid dynamic (CFD) systems at the corresponding characteristic scales can be included in the loss function, forming the PDE-constrained optimization problem \cite{de2015numerical}.
Such a methodology, called end-to-end training, is in favor as it covers the dynamics of a continuous-time model with necessary prior knowledge and physical structure.
Each moment of the training data and predictions can be aligned and both supervised and unsupervised learning can be adapted.

Due to the high computational cost of CFD solvers, gradient-based optimization is arguably a natural pair with end-to-end training.
However, the existence of solution trajectories of governing equations has made it trickier to obtain accurate gradient information.
As analyzed in \cite{johnson2012notes}, direct evaluation of gradients in initial-value problems for partial differential equations can lead to an explosion of computational complexity, making it impractical to perform for a large number of flow variables or control parameters.
The adjoint method addresses this challenge by introducing auxiliary variables and constructing the dual form of the optimization problem, in which the vector-Jacobian products replace the costly evaluations of Jacobian \cite{strang2007computational,givoli2021tutorial}.
It is a well-suited tool for optimization and sensitivity analysis in high-dimensional space, e.g., aerodynamic shape design and optimization \cite{jameson2003aerodynamic}.
Note that the analog of the adjoint method in deep learning is the reverse-mode automatic differentiation (AD), i.e., what we know as backpropagation, where the chain rule is applied as a sequence of vector-Jacobian product operations from the loss function.

Differentiable programming (denoted as $\partial P$) has become a prominent notion for conducting scientific machine learning research.
Unlike the convention where AD is limited to accumulating the gradients of matrix operations, neural networks in $\partial P$ are regarded as generic nonlinear functions specified through a computer program.
Other differentiable physical models or agents, including differential equations, can be incorporated as nodes in a computation graph equivalent to a neural network.
This has allowed us to integrate principled differentiable operations and have them act as building blocks for each other to better approximate the structure of the problem in the task at hand, e.g., neural ODE models \cite{chen2018neural} and parametric high-dimensional differential equations \cite{rackauckas2020universal}.
The differentiability of the individual nodes in the computation graph facilitates taking gradients of the computer program under the chain rule, which is the essential difference between $\partial P$ and classical computer programming \cite{blondel2024elements}.
In summary, $\partial P$ provides a novel paradigm that unifies data structures and control flows to enable end-to-end AD and gradient-based optimization in machine learning and scientific computing tasks.

Although AD undoubtedly revolutionizes the paradigm of computing the gradient of complicated functions, which can be extremely tedious (or even impossible) to implement manually, it is important to note that it is not a panacea.
The practice of AD faces at least two challenges.
First, existing AD engines often restrict types and styles of code.
In the case of JAX \cite{jax2018github}, for example, all functions need to be mathematically valid, (a.k.a. pure functions), and thus control flows must be organized through functional programming.
Such requirements are not trivial to meet, especially when external libraries from other compilers or languages are called, or heterogeneous computing is used.
Second, AD may generate less efficient codes.
An example goes to iterative schemes, e.g., the Krylov subspace methods \cite{liesen2013krylov}.
Here, the direct application of AD to the solution of a linear system leads to a high computational overhead since the AD engine will treat the iteration as recurrence and store all intermediate steps.
A feasible workaround for the above challenges is to implement the gradient manually, e.g., by applying the adjoint method to the final-state solution, and to compose the self-defined vector-Jacobian product into the chain rule for the surrounding operations.
In other words, AD can be beneficial, but the power of $\partial P$ can only be maximized through sensible human intervention (designing differentiable operations and combining them with efficient adjoints).

There is an emerging consensus in the academic community on the importance of $\partial P$ in CFD practices.
Among others, Belbute-Peres et al. combined a differentiable CFD simulator and graph neural networks to accelerate the CFD prediction \cite{belbute2020combining}.
Zhuang et al. built a set of differentiable codes to learn the optimal discretization for passive scalar advection in turbulent flows \cite{zhuang2021learned}.
Bezgin et al. constructed a differentiable CFD program for multi-phase flows based on the JAX engine \cite{bezgin2023jax}.
Fan and Wang employed $\partial P$ to model fluid-structure interaction efficiently \cite{fan2024differentiable}.
Ho and Farhat developed a differentiable embedded boundary method for the sake of aerodynamic optimization \cite{ho2023aerodynamic}.
Kochkov et al. employed end-to-end differentiable learning for subgrid model discovery in turbulence \cite{kochkov2021machine}.
Um et al. placed differentiable physics into the training process to reduce the error of iterative PDE solvers.
To the best of the author's knowledge, the existing work has focused on solving single governing equations, while work on the physics of multi-scale flows with multiple governing equations and multiple degrees of freedom is limited.

This paper serves as an exploration of the use of $\partial P$ to solve continuum and rarefied flows.
Based on the kinetic theory of gases, the parametric kinetic model and solution algorithm with differentiable operations are built for the Boltzmann equation and its hydrodynamic asymptotics.
The continuous adjoint equations are developed based on the semi-discrete governing equations derived from the finite volume method and then bundled into the AD engine.
The design parameters in flow models and numerical methods, especially in neural networks, can then be optimized on the fly with the gradients through the backward passes of the whole simulation program.
Thus, a unified differentiable simulator is constructed that can tightly integrate the solution and optimization processes and is suitable for forward and inverse problems arising in rarefied and multi-scale gaseous flows.
The program implementation is based on the Julia language, which enables language-wide AD via source-to-source transformation \cite{innes2019differentiable,moses2020instead}.
Numerical experiments for both continuum and rarefied flow problems will be presented to elucidate the $\partial P$-based solution paradigm and validate the computer program.
For reproducible science, the relevant codes (augmented by Kinetic.jl, an indigenously developed differentiable framework designed for scientific and neural computing tasks \cite{xiao2021kinetic}) for this paper are available under the MIT license \footnote{\url{https://github.com/vavrines/KitAD.jl}}.

The rest of this paper is organized as follows. 
Section \ref{sec:theory} presents a brief introduction to the kinetic theory of gases and numerical discretizations.
Section \ref{sec:adjoint} derives the adjoint equations and illustrates the joint use with AD in flow optimization problems.
Section \ref{sec:algorithm} describes the complete solution and optimization algorithms.
Section \ref{sec:experiment} includes numerical experiments to demonstrate the validity and performance of the current method. 
The last section is the conclusion.
The nomenclature of this paper is presented in Table \ref{table:nomenclature}.

\section{Basic Theory}\label{sec:theory}

The kinetic theory can inscribe the flow physics of rarefied and continuum gases.
Lying at its core, the fluid is modeled as a many-particle system and its time-space evolution is statistically tracked using the single-particle distribution function.
In the absence of internal degrees of freedom and external force, the Boltzmann equation for the distribution function $f(t,\mathbf x,\mathbf v)$ writes
\begin{equation}
    \frac{\partial f}{\partial t} + \mathbf v \cdot \nabla_\mathbf x f = \mathcal Q(f,f) = \int_{\mathbb{R}^{3}} \int_{\mathbb S^{2}}\left[f\left(\mathbf{v}^{\prime}\right) f\left(\mathbf{v}_{*}^{\prime}\right)-f(\mathbf{v}) f\left(\mathbf{v}_{*}\right)\right] \mathcal{K}(\cos \theta, g) d \boldsymbol\beta d \mathbf{v}_{*},
\label{eq:boltz}
\end{equation}
where $\{\mathbf v, \mathbf v_*\}$ and $\{\mathbf v', \mathbf v_*'\}$ denote the pre- and post-collision velocities of two classes of colliding particles.
The collision kernel $\mathcal{K}(\cos \theta, g)$ is a measure of the probability of collisions in different directions, where $\theta$ is the deflection angle and $g = |\mathbf g| = |\mathbf v - \mathbf v_*|$ is the magnitude of relative pre-collision velocity.
The deflection angle satisfies the relation $\theta=\boldsymbol\beta \cdot \mathbf g / g$, where the solid angle $\boldsymbol\beta$ is the unit vector along the relative post-collision velocity $\mathbf v' - \mathbf v_*'$.

The Boltzmann equation is an integro-differential equation with extremely high dimensionality and nonlinearity.
To reduce the computational overhead of the fivefold integral, simplified relaxation models, e.g. the Bhatnagar-Gross-Krook (BGK) model, are commonly adopted in the simulation of complex flows.
The relaxation model writes
\begin{equation}
    \frac{\partial f}{\partial t} + \mathbf v \cdot \nabla_\mathbf x f = \mathcal R(f) = \nu (\mathcal E - f),
\label{eq:bgk}
\end{equation}
where $\mathcal E$ is the equilibrium distribution of relaxation directions and $\nu$ denotes the relaxation frequency.
In the BGK model, $\mathcal E$ takes the form of the Maxwellian, i.e.,
\begin{equation}
    \mathcal E = \mathcal M := \rho\left(\frac{m}{2\pi k T}\right)^{3/2} \exp(-\frac{m}{2kT} \left(\mathbf v - \mathbf V)^2 \right),
\end{equation}
where $\{\rho,\mathbf V,T\}$ are the macroscopic density, velocity and temperature, $m$ is the molecular mass, $k$ is the Boltzmann constant.
The kinetic equations provide a mesoscopic view to describe particle transports and are consistent with first physical laws, including boundedness, conservation, invariance, and entropy principle \cite{alldredge2019regularized,xiao2023relaxnet}.
In the following, we denote the Boltzmann collision operator $\mathcal Q(f,f)$ and relaxation term $\mathcal R(f)$ uniformly as $\mathcal Q(f)$.

A particle distribution function is related to a unique macroscopic state.
The conservative variables in fluid mechanics can be obtained by taking moments of particle distribution function over velocity space, i.e.,
\begin{equation}
    \mathbf{W}(t, \mathbf{x})=\left(\begin{array}{c}
    \rho \\
    \rho \mathbf{V} \\
    \rho E
    \end{array}\right) := \int_{\mathbb R^3} f \psi d \mathbf{v},
\end{equation}
where $\psi=(1,\mathbf v,\mathbf v^2/2)^T$ is a vector of collision invariants satisfying $\int_{\mathbb R^3} \mathcal Q(f) \psi d \mathbf v = 0$, and temperature is defined as
\begin{equation}
    \frac{3}{2} kT = \frac{1}{2n} \int_{\mathbb R^3} (\mathbf v - \mathbf V)^2 f d\mathbf v,
\end{equation}
where $n$ is the number density of gas.
Taking conservative moments of the kinetic equation (\ref{eq:boltz}) or (\ref{eq:bgk}) yields the conservation laws which write
\begin{equation}
    \partial_t \mathbf W + \int_{\mathbb R^3} \psi \mathbf v \cdot \nabla_{\mathbf x} f d\mathbf v=0,
\end{equation}
i.e.,
\begin{equation}
\begin{aligned}
    &\frac{\partial \rho}{\partial t} + \nabla_{\mathbf x} \cdot (\rho \mathbf V)=0, \\
    &\frac{\partial \rho \mathbf V}{\partial t} + \nabla_{\mathbf x} \cdot (\rho \mathbf V \otimes \mathbf V)-\nabla_{\mathbf x} \cdot \mathbf P=0, \\
    &\frac{\partial \rho E}{\partial t} + \nabla_{\mathbf x} \cdot (\rho E\mathbf V)-\nabla_{\mathbf x} \cdot (\mathbf P \cdot \mathbf V)+\nabla_{\mathbf x} \cdot \mathbf q=0,
\end{aligned}
\label{eq:conservation laws}
\end{equation}
where $\otimes$ denotes dyadic product, and the stress tensor $\mathbf P$ and heat flux $\mathbf q$ are defined as
\begin{equation}
    \mathbf P=\int_{\mathbb R^3} (\mathbf v-\mathbf V)\otimes(\mathbf v-\mathbf V)fd\mathbf v, \quad \mathbf q=\int_{\mathbb R^3} \frac{1}{2}(\mathbf v-\mathbf V)(\mathbf v-\mathbf V)^2fd\mathbf v.
\end{equation}
Choosing a suitable closure strategy for $\mathbf P$ and $\mathbf q$ yields solvable Euler, Navier-Stokes, and extended hydrodynamic equations \cite{torrilhon2016modeling}.

CFD is dedicated to approximating the solution of governing equations at a discrete level.
For Eq.(\ref{eq:bgk}), we consider the domain $\mathbf \Omega=\mathbf \Omega_{\mathbf x} \times \mathbf \Omega_{\mathbf v}$ with $N_{x}\times N_{v}$ non-overlapping cells,
\begin{equation}
\begin{aligned}
    &\mathbf{\Omega}_{\mathbf x} = \bigcup_{i=1}^{N_{x}} \mathbf \Omega_{i}, \quad \bigcap_{i=1}^{N_{x}} \mathbf \Omega_{i} = \emptyset, \\
    &\mathbf{\Omega}_{\mathbf v} = \bigcup_{j=1}^{N_{v}} \mathbf \Omega_{j}, \quad \bigcap_{j=1}^{N_{v}} \mathbf \Omega_{j} = \emptyset,
\end{aligned}
\end{equation}
and the particle distribution function is approximated as
\begin{equation}
    f \simeq \bigoplus_{i=1,j=1}^{N_{x},N_{v}} f_{i,j},
\end{equation}
where $f_{i,j}$ denotes the piecewise-defined distribution function inside each cell.
Different discretization methods can be used to approximate the solution $f_{i,j}$.
Here we take the finite volume method as an example to illustrate.

We define the cell-averaged distribution function as
\begin{equation}
    \bar f_{i,j}=\frac{1}{V_i V_j} \int_{\mathbf \Omega_{i}}\int_{\mathbf \Omega_{j}} f(t,\mathbf x,\mathbf v) d\mathbf v d\mathbf x,
\end{equation}
where $V_{i}$ and $V_{j}$ are the volumes of $\mathbf \Omega_{i}$ and $\mathbf \Omega_{j}$, respectively.
Integrating Eq.(\ref{eq:bgk}) with respect to $\mathbf x$ and applying Gauss's law yields 
\begin{equation}
\begin{aligned}
    \frac{\partial \bar f_{i,j}}{\partial t} &= -\frac{1}{V_{i}} \oint_{\partial \mathbf \Omega_i} \mathbf F^f_j(t,\mathbf x) \cdot d\mathbf S + \mathcal Q(\bar f_{i,j}) \\
    &= -\frac{1}{V_{i}} \sum_{k=1}^{N_f} \mathbf F_{k,j}^f \cdot \Delta \mathbf S_k + \bar \nu_i (\bar{\mathcal E}_{i,j} - \bar f_{i,j}),
\end{aligned}
\label{eq:fvm micro}
\end{equation}
where $\mathbf F^f$ denotes the numerical flux of distribution function, $\mathbf S=\mathbf n \Delta S$ is the area vector pointing out of the cell, and $N_f$ is the number of faces.
Different approaches can be employed to construct the numerical flux $\mathbf F^f$.
Since the kinetic equation is consistent with particle transport processes, a neat choice is to build the numerical flux in an upwind manner.
We take the $k$-th face of cell $i$ as an example and assume that the cell index on the other side of the face is $i+1$, then the numerical flux is constructed as
\begin{equation}
\begin{aligned}
    &\mathbf F^f_{k,j} = \mathbf v_j f^f_{k,j}, \\
    &f_{k,j}^f = \hat f_{i,j}^k H(\mathbf n \cdot \mathbf v_j) + \hat f_{i+1,j}^k (1-H(\mathbf n \cdot \mathbf v_j)),
\end{aligned}
\label{eq:flux}
\end{equation}
where $\hat f_{i,j}^k$ denotes the reconstructed distribution function at the face based on in-cell slopes, and $H$ is the Heaviside step function.

Following the derivation of Eq.(\ref{eq:conservation laws}), taking moments of Eq.(\ref{eq:fvm micro}) over velocity space $\mathbf \Omega_{\mathbf v}$ yileds the semi-discrete formulation of conservation laws, i.e.,
\begin{equation}
\begin{aligned}
    \frac{\partial \bar{\mathbf W}_{i}}{\partial t} &= -\frac{1}{V_{i}} \oint_{\partial \mathbf \Omega_i} \mathbf F^W \cdot d\mathbf S = -\frac{1}{V_{i}} \sum_{k=1}^{N_f} \mathbf F_{k}^W \cdot \Delta \mathbf S_k.
\end{aligned}
\label{eq:fvm macro}
\end{equation}
Here, the cell-averaged conservative variables in $\mathbf \Omega_{i}$ can be approximated by numerical quadrature at the discrete level, i.e.,
\begin{equation}
    \bar{\mathbf W}_{i} := \int_{\mathbb R^3} f_{i} \psi d \mathbf{v} \simeq \sum_{j=1}^{N_{v}} w_j f_{i,j} \psi_j,
\end{equation}
where $w_j$ denotes the quadrature weights.

Given the number of elements $N_{x}$ and $N_{v}$, Eq.(\ref{eq:fvm micro}) and (\ref{eq:fvm macro}) form a system of ordinary differential equations (ODEs) or differential-algebraic equations (DAEs), respectively.
We uniformly denote the variables as $\mathbf u \in \mathbb R^{N_{u}}$, and the solution system can then be written as
\begin{equation}
\begin{aligned}
    &\frac{\partial \mathbf u}{\partial t}=\mathcal F(t,\mathbf u,\mathbf p), \\
    &\mathcal C(t,\mathbf u,\mathbf p)=0,
\end{aligned}
\label{eq:ode}
\end{equation}
where $\mathbf p\in\mathbb R^{N_{p}}$ is the collection of control parameters of the solution algorithm.
The contributions of numerical fluxes and source terms are represented by the operator $\mathcal F$, and the initial and boundary conditions are bounded by the operator $\mathcal C$.

The solution of Eq.(\ref{eq:ode}) can be obtained by integrating it along the time direction.
Note that the relaxation frequency in Eq.(\ref{eq:bgk}) is proportional to the gas density, and thus the choice of the integrator is related to the regime of the flow problem.
In the continuum limit, Eq.(\ref{eq:bgk}) can become stiff.
Therefore, an appropriate integrator is chosen in the hope that it is efficient and A- or L-stable for stiff and oscillatory problems.
The choices available include the backward differentiation formula (BDF) \cite{brayton1972new}, multi-stage implicit Runge–Kutta (IRK) methods \cite{jameson2017evaluation}, and implicit-explicit (IMEX) methods \cite{ascher1995implicit}.
The performance of different integrators for solving kinetic equations is briefly summarized in \cite{xiao2021flux}.





\section{Differentiation Strategy}\label{sec:differentiation}

\subsection{Adjoint System}\label{sec:adjoint}

For the differential-equation-constrained optimization problem, a cost function denoted $C(\mathbf u,\mathbf p)$ will be computed throughout the solution trajectory of the governing equation.
This problem can often be handled efficiently by the adjoint sensitivity method \cite{errico1997adjoint}, which is well-suited for situations requiring the sensitivity analysis of a scalar (or low-dimensional) function of the solution with respect to a potentially large number of parameters.
We follow the derivation presented in \cite{cao2003adjoint}, but modify it to specialize on the adjoint system of Eq.(\ref{eq:ode}).
Eq.(\ref{eq:ode}) is index-0 and index-1 differential-algebraic equations (DAEs) for hydrodynamic and kinetic equations.
Since it is linear with respect to the derivative term, we introduce the linear mass matrix and reformulate it as
\begin{equation}
    M \mathbf u' = \mathcal G(t,\mathbf u,\mathbf p),
\end{equation}
where $\mathbf u'$ denotes the time derivative for brevity.

For a time-varying problem, a viable cost function can be constructed as
\begin{equation}
    C(\mathbf u,\mathbf p)=\int_{t_0}^{t_1} g(t,\mathbf u,\mathbf p)dt,
\label{eq:continuous cost}
\end{equation}
where $t_0$ and $t_1$ denote two moments in time.
We expect to obtain the derivative $\partial C/\partial \mathbf p$, and the problem translates into computing the intermediate quantity $\lambda$ (called the adjoint variable) as the solution of the adjoint system.
The derivatives $\partial_{\mathbf u} C$ and $\partial_{\mathbf p} C$ should exist and be bounded.
We introduce the adjoint variable $\lambda$ as a Lagrange multiplier that conforms
\begin{equation}
    I(\mathbf u, \mathbf p)=C(\mathbf u, \mathbf p)-\int_{t_0}^{t_1} \lambda^* \mathcal H(t,\mathbf u,\mathbf u',\mathbf p) d t ,
\end{equation}
where $\lambda^*$ denotes the conjugate transpose of $\lambda$, and $\mathcal H = M \mathbf u' - \mathcal G=0$.
The partial derivatives of $C$ with respect to $\mathbf p$ can thus be written as
\begin{equation}
    \frac{\partial C}{\partial \mathbf p}=\frac{\partial  I}{\partial  \mathbf p}=\int_{t_0}^{t_1}\left(g_{\mathbf p}+g_{\mathbf u} \mathbf u_{\mathbf p}\right) d t-\int_{t_0}^{t_1} \lambda^*\left(\mathcal H_{\mathbf p}+\mathcal H_{\mathbf u} {\mathbf u}_{\mathbf p}+\mathcal H_{\mathbf u'} {\mathbf u}'_{\mathbf p}\right) d t.
\end{equation}
Applying integration by parts leads to
\begin{equation}
    \frac{\partial C}{\partial \mathbf p}=\int_{t_0}^{t_1}\left(g_{\mathbf p}-\lambda^*\mathcal H_{\mathbf p}\right) d t
    +\int_{t_0}^{t_1} (g_{\mathbf u}-\lambda^* \mathcal H_{\mathbf u}+(\lambda^* \mathcal H_{\mathbf u'})') \mathbf u_{\mathbf p} d t 
    - [\lambda^* \mathcal H_{\mathbf u'} \mathbf u_{\mathbf p}]_{t_0}^{t_1} .
\end{equation}
We require that
\begin{equation}
    g_{\mathbf u}-\lambda^* \mathcal H_{\mathbf u}+(\lambda^* \mathcal H_{\mathbf u'})'=0,
\end{equation}
and
\begin{equation}
    \lambda^* \mathcal H_{\mathbf u'} \vert_{t=t_1} =0,
\end{equation}
and thus the sensitivity equation for $\partial C/\partial \mathbf p$ becomes
\begin{equation}
\begin{aligned}
    \frac{\partial C}{\partial \mathbf p}&=\int_{t_0}^{t_1} \left(g_{\mathbf p}-\lambda^* \mathcal H_{\mathbf p}\right) d t +
    (\lambda^* \mathcal H_{\mathbf u'} \mathbf u_{\mathbf p}) \vert_{t=t_0} \\
    &=\int_{t_0}^{t_1} \left(g_{\mathbf p}+\lambda^* \mathcal G_{\mathbf p}\right) d t + \lambda^*(t_0) M \mathbf u_{\mathbf p}.
\end{aligned}
\label{eq:sensitivity}
\end{equation}
Thus we have derived the sensitivity equation along with the adjoint DAE system for $\lambda$ and its boundary condition
The derivative of the solution with respect to a cost function can be obtained by solving the adjoint and sensitivity equations in turn.
Note that even if $C$ is discrete, it can be similarly expressed as
\begin{equation}
    C(\mathbf u,\mathbf p) = \int_{t_0}^{t_1} \sum^{N_{d}}_i \|\mathbf d_i-\mathbf u(t_i,\cdot)\|^2 \delta(t_i - t) dt,
\label{eq:discrete cost}
\end{equation}
in which case
\begin{equation}
    g_{\mathbf u}(t_i)=2(\mathbf d_i-\mathbf u(t_i,\cdot)),
\end{equation}
where $\mathbf d_i$ denotes the data point at $t_i$ \cite{rackauckas2020universal}.
The same steps can then be applied subsequently.




\subsection{Automatic Differentiation}\label{sec:ad}

In $\partial P$, the solution of the adjoint system is nested within the AD workflow.
We consider a computer program in which a numerical operation $\digamma: \mathcal S_0 \rightarrow \mathcal S_K$ can generally be written as a sequence of compositions, i.e.,
\begin{equation}
    \digamma = \digamma_K \circ \digamma_{K-1} \circ \cdots \circ \digamma_1,
\label{eq:chain}
\end{equation}
where $\digamma_k: \mathcal S_{k-1} \rightarrow \mathcal S_k$.
The inputs and outputs of functions are $\mathbf{s}_{k-1} \in \mathcal S_{k-1}$ and $\mathbf{s}_{k} \in \mathcal S_{k}$, respectively.
Note that multiple dependencies of intermediate functions can be efficiently represented using directed acyclic graphs (DAGs), thus keeping the consistency of the above equation.
Based on the chain defined in Eq.(\ref{eq:chain}), the full Jacobian matrix can be obtained as
\begin{equation}
    \partial \digamma(\mathbf s_0)=\partial \digamma_K(\mathbf s_{K-1}) \partial \digamma_{K-1}(\mathbf s_{K-2}) \cdots  \partial\digamma_{1}(\mathbf s_{0}).
\label{eq:jacobian chain}
\end{equation}
The computational overhead of the above equation is high due to the matrix multiplications of intermediate Jacobians.
However, in most cases, we need the derivatives of the composition of $\digamma$ and a scalar-valued cost function $C \circ \digamma$.
This translates into solving the right or left multiplication of the Jacobian, rather than itself.
Forward-mode and reverse-mode ADs are developed on this basis, respectively.

\vspace{4mm}
\noindent\textbf{Forward-mode AD}

\begin{figure}
    \centering
    \includegraphics[width=\textwidth]{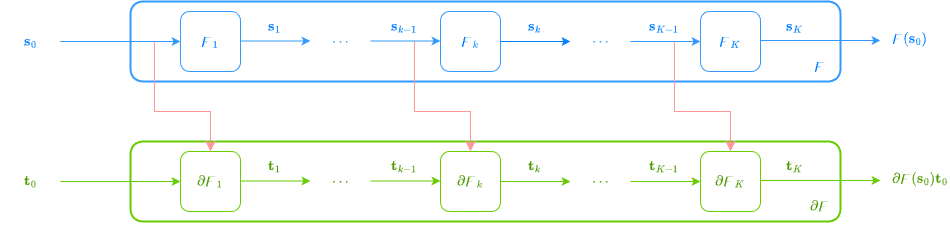}
    \caption{Schematic of forward-mode automatic differentiation for a sequence of functions.}
    \label{fig:forward}
\end{figure}

The computation of Jacobian can be understood as a composition of primitively known linear maps, i.e.,
\begin{equation}
    \partial \digamma(\mathbf s_0)=\partial \digamma_K(\mathbf s_{K-1}) \circ \partial \digamma_{K-1}(\mathbf s_{K-2}) \circ \cdots \circ \partial\digamma_{1}(\mathbf s_{0}).
\end{equation}
The evaluation of $\partial\digamma(\mathbf s_0)$ on an input vector $\mathbf w$ can be performed by computing Jacobian-vector products (JVPs) along the same direction as the computation of intermediate states $\mathbf s_k$, hence the name forward-mode AD.
This corresponds to the right multiplication of Eq.(\ref{eq:jacobian chain}).
Such a scheme can often be succinctly implemented using dual numbers \cite{revels2016forward}.
Since the computational complexity and memory load of computing $\partial \digamma_k$ is comparable to the cost of computing $\digamma_k$, the computational cost of a JVP is roughly twice that of $\digamma_k$.
The schematic of forward-mode AD is presented in Figure \ref{fig:forward} and the detailed solution steps can be found in Algorithm \ref{alg:forward}.

\begin{algorithm}
\caption{Forward-mode automatic differentiation for a sequence of functions}
\begin{algorithmic}
    \State \textbf{Function:} $\digamma = \digamma_K \circ \digamma_{K-1} \circ \cdots \circ \digamma_1$
    \State \textbf{Input variable:} $\mathbf s_0 \in \mathcal S_0$
    \State \textbf{Input direction:} $\mathbf w \in \mathcal S_0$
    \State {Initialize} $\mathbf t_0 = \mathbf w$
    \For {$k=1,\dots, K$}
        \State {Compute} $\mathbf s_k = \digamma_k(\mathbf s_{k-1})$
        \State {Compute} $\mathbf t_k = \partial\digamma_k(\mathbf s_{k-1}) \mathbf t_{k-1}$
    \EndFor
    \State \textbf{Output function value:} $\digamma(\mathbf s_0)=\mathbf s_K$
    \State \textbf{Output JVP:} $\partial\digamma(\mathbf s_0)\mathbf w=\mathbf t_K$
\end{algorithmic}
\label{alg:forward}
\end{algorithm}

\vspace{5mm}
\noindent\textbf{Reverse-mode AD}

\begin{figure}
    \centering
    \includegraphics[width=\textwidth]{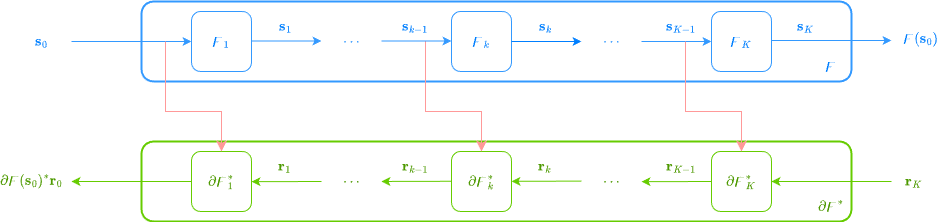}
    \caption{Schematic of reverse-mode automatic differentiation for a sequence of functions.}
    \label{fig:reverse}
\end{figure}

The gradient of $C \circ \digamma$ takes the form
\begin{equation}
    \nabla(C \circ \digamma)(\mathbf s_0)=\partial \digamma(\mathbf s_0)^* \nabla C(\digamma(\mathbf s_0)),
\end{equation}
where the adjoint map is defined as $\partial \digamma(\mathbf s_0)^* : \mathcal S_K \rightarrow \mathcal S_0$,
and it yields
\begin{equation}
    \partial \digamma(\mathbf s_0)^*=\partial\digamma_{1}(\mathbf s_{0})^* \circ \partial\digamma_{2}(\mathbf s_{1})^* \circ \cdots \circ \partial \digamma_K(\mathbf s_{K-1})^*.
\end{equation}
Each intermediate adjoint $\partial \digamma_k(\mathbf s_{k-1})^*$ is equivalent to a vector-Jacobian product (VJP).
Here, VJPs are computed recursively along the opposite direction of $\mathbf s_k$, hence the name reverse-mode AD.
The computational complexity of VJPs, like JVPs, is roughly twice that of the original function.
The memory usage grows linearly with respect to the sequence length $K$.
The schematic of reverse-mode AD is presented in Figure \ref{fig:reverse} and the detailed solution steps can be found in Algorithm \ref{alg:reverse}.

\begin{algorithm}
\caption{Reverse-mode automatic differentiation for a sequence of functions}
\begin{algorithmic}
    \State \textbf{Function:} $\digamma = \digamma_K \circ \digamma_{K-1} \circ \cdots \circ \digamma_1$
    \State \textbf{Input variable:} $\mathbf s_0 \in \mathcal S_0$
    \State \textbf{Output direction:} $\mathbf w \in \mathcal S_K$
    \For {$k=1,\dots, K$}\Comment{Forward pass}
        \State {Compute} $\mathbf s_k = \digamma_k(\mathbf s_{k-1})$
    \EndFor
    \State {Initialize} $\mathbf r_K = \mathbf w$
    \For {$k=1,\dots, K$}\Comment{Backward pass}
        \State {Compute} $\mathbf r_{k-1} = \partial\digamma_k(\mathbf s_{k-1})^* \mathbf r_{k}$
    \EndFor
    \State \textbf{Output function value:} $\digamma(\mathbf s_0)=\mathbf s_K$
    \State \textbf{Output VJP:} $\partial\digamma(\mathbf s_0)^*\mathbf w=\mathbf r_0$
\end{algorithmic}
\label{alg:reverse}
\end{algorithm}

Note that the JVP and VJP operations can be generalized within the framework of differential geometry based on the definition of directional derivatives, where the JVP corresponds to the pushforward operator acting on tangent vectors, while the VJP amounts to the pullback operator on cotangent vectors.
The pushforward and pullback operations can be decomposed similarly according to the defined chain rules.

As discussed in \cite{blondel2024elements}, the computational efficiency of forward- and reverse-mode ADs depends on the dimension of $\mathcal S_k$.
Given $\mathcal S_k \subseteq \mathbb R^{D_k}$, the forward-mode AD is more advantageous in the case of $D_K\geq D_0$, while the reverse-mode is more favorable when $D_K<D_0$.
Note that the latter is the more common case when a considerable number of parameters are involved, as is the case in neural networks.

Existing AD implementations are mainly divided into several categories.
One classical option is tape-based AD, which leverages a data structure (tape) to record the sequence of operations \cite{paszke2019pytorch}.
Another AD approach is source-to-source transformation, where the derivatives are generated analytically through code generation \cite{van2018automatic}.
In the current work, we employ Enzyme, an AD engine that performs code generation at the intermediate representation (IR) level of the LLVM  \cite{moses2020instead}.
Based on the predefined chain rules \cite{white2024chain}, the adjoint system in Section \ref{sec:adjoint} will be automatically invoked when the solution of differential equations is encountered during the differentiation process.
Therefore, the adjoint and AD systems can be bundled and work together as a whole.

\section{Solution Algorithm}\label{sec:algorithm}

\subsection{Machine Learning}

Machine learning models are parametric representations that map inputs (features) and outputs (targets) without being explicitly programmed.
Among these, neural networks are the most dominant model in the current wave of deep learning.
Taking the feedforward neural network $\mathrm{NN}_{\boldsymbol\theta}$, as an example, it can be viewed as a sequence of parameterized functions, i.e.,
\begin{equation}
\begin{aligned}
    \mathbf s_0 & :=\mathbf{u},\\
    \mathbf s_1 & :=\mathcal L_1\left(\mathbf s_0, \boldsymbol{\theta}_1\right), \\
    \mathbf s_2 & :=\mathcal L_2\left(\mathbf s_1, \boldsymbol{\theta}_2\right), \\
    & \vdots \\
    \mathbf {s}_K & :=\mathcal L_K\left(\mathbf{s}_{K-1}, \boldsymbol{\theta}_K\right),
\end{aligned}
\label{eq:nn}
\end{equation}
where $\mathcal L_k$ indicates a function layer, and $\boldsymbol{\theta}:=(\boldsymbol\theta_1,\dots,\boldsymbol\theta_K)$ denotes the parameters to be optimized.
As a typical parameterization, the multi-layer perceptron (MLP) adopts the fully-connected layers of the form
\begin{equation}
    \mathcal L_k := \phi_k (\boldsymbol \omega_k \mathbf s_{k-1} + \mathbf b_k)
\label{eq:perceptron}
\end{equation}
where the affine layer with the weight matrix $\boldsymbol{\omega}_k$ and bias vector $\mathbf b_k$ and the activation function $\phi_k$ are combined to describe a
nonlinear transformation.
The affine layer in Eq.(\ref{eq:perceptron}) can be replaced by other linear functions, e.g., convolution and filtering, while Eq.(\ref{eq:nn}) can be replaced with more general models with more complex structures, e.g., that used in residual and recurrent learning.

Machine learning provides versatile means for describing non-equilibrium flows.
Parameterized models can effectively promote the applicability of constitutive relations, numerical fluxes, source terms, and related components.
Taking neural networks as an example, based on Eq.(\ref{eq:ode}), a unified mechanical-neural model can be formulated as
\begin{equation}
\begin{aligned}
    &\frac{\partial \mathbf u}{\partial t}=\mathcal F(t, \mathbf u, \boldsymbol\alpha, \mathrm{NN}_{\boldsymbol\theta}(t,\mathbf u)), \\
    &\mathcal C(t,\mathbf u,\boldsymbol\alpha,\boldsymbol\theta)=0,
\end{aligned}
\label{eq:ude}
\end{equation}
where $\mathrm{NN}_{\boldsymbol\theta}$ denotes the forward pass of a neural network, $\mathbf p=(\boldsymbol\alpha,\boldsymbol\theta)$ signifies the parameters in the mechanical and neural network models, respectively.
The architecture of Eq.(\ref{eq:ude}) is similar to that of neural ordinary differential equations, and thus offers advantages, e.g., memory efficiency and adaptive computation \cite{chen2018neural}.
As a trainable system, it has the same solution methodology as Eq.(\ref{eq:ode}).
With the introduction of neural networks, the dimensionality of the model's parameter space increases dramatically, and the ability to depict non-equilibrium flows can be subsequently improved.


\subsection{Solution Algorithm}

The unified mechanical-neural model developed in Eq.(\ref{eq:ude}) requires the solution of both forward and optimization problems.
The solution of the forward problem follows a similar principle as Eq.(\ref{eq:ode}) in Section \ref{sec:theory}, where an appropriate integrator implemented with differentiable operations is employed to iterate numerical solutions.
For the constrained optimization problem, a cost function is needed to align the numerical solution towards the referenced data points.
A commonly adopted definition of the cost function for supervised learning tasks writes
\begin{equation}
    C = \sum_i^{N_c} \sum^{N_{t}}_j \|\mathbf u_i^{\mathrm{ref}}(t_j)-\mathbf u_i(t_j)\|^2 + \epsilon \|\boldsymbol{\theta}\|^2,
\label{eq:cost}
\end{equation}
which corresponds to the discrete cost function defined in Eq.(\ref{eq:discrete cost}).
Here, $N_c$ represents the number of different flow conditions to be simulated and $N_t$ is the number of time steps.
Thus, the trajectories of the numerical solution are tracked upon the total number of data samples $N_{d}=N_c N_t$.
The referenced solution $\mathbf u^{\mathrm{ref}}$ can be obtained from fine-grained models with high confidence and fidelity, e.g., molecular simulation results.
The $L_2$ regularization term mitigates overfitting and improves model generalization by penalizing large weights.
The regularization parameter $\epsilon$ is an empirical parameter that needs to be chosen in a trade-off between bias and variance.

\begin{figure}
    \centering
    \includegraphics[width=\textwidth]{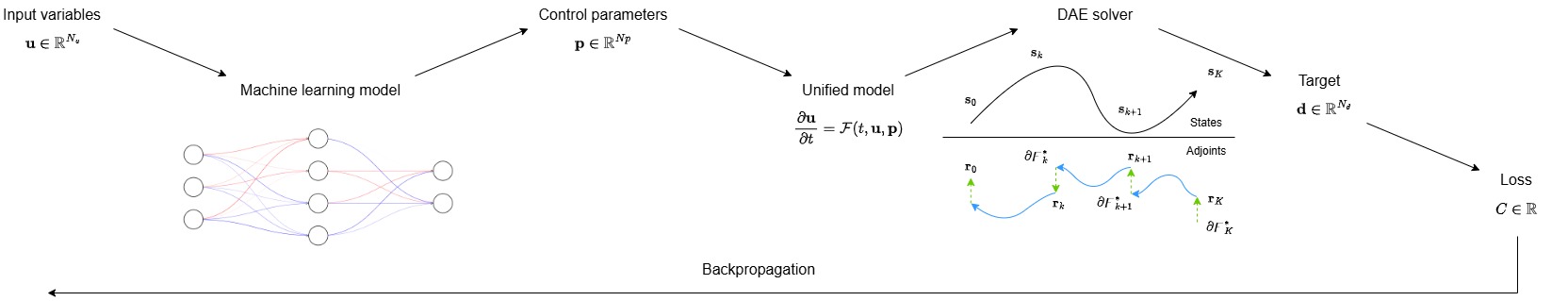}
    \caption{Flow of the solution and training algorithms for the unified mechanical-neural model based on differentiable programming.}
    \label{fig:flowchart}
\end{figure}

The gradient information of the cost function $C$ is required to leverage gradient-based optimization methods.
Since a considerable number of parameters is introduced with neural networks, usually the reverse-mode AD is more favored.
As discussed in Section \ref{sec:ad}, the solution algorithm can be expressed as a sequence of operations, and its derivatives can be computed with the help of sequenced VJPs, which allows for a recursive decomposition of a primitively known set of pullbacks.
The predefined adjoints can be incorporated into the chain rule to accelerate the gradient computation.
After the derivatives of the cost function have been obtained, gradient-descent-type methods can be employed to optimize the cost function efficiently, e.g. the first-order stochastic gradient descent (SGD) method \cite{bottou2010large}, and the second-order Broyden–Fletcher–Goldfarb–Shanno (BFGS) method \cite{liu1989limited}.
The optimized mechanical and neural parameters are then used to perform the subsequent forward computation for the unified model, and so on iteratively.
The flow of the $\partial P$-based solution algorithm is briefly summarized in Figure \ref{fig:flowchart}.

\section{Numerical Experiments}\label{sec:experiment}

In this section, we will conduct numerical experiments to validate the $\partial P$-based solution algorithm.
To illustrate the applicability of the methodology to cross-scale flows, cases with different degrees of gas rarefaction are considered.
Dimensionless variables are uniformly adopted in the numerical simulations, which are defined as
\begin{equation}
\begin{aligned}
    &\tilde t = \frac{t}{L_0/V_0}, \ \tilde{\mathbf x} = \frac{\mathbf x}{L_0}, \ \tilde{\rho} = \frac{\rho}{\rho_0}, \ \tilde{\mathbf V} = \frac{\mathbf V}{V_0}, \ \tilde{T} = \frac{T}{T_0}, \ \tilde{E} = \frac{E}{V_0^2}, \ \\
    &\tilde{\mathbf P} = \frac{\mathbf P}{\rho_0 V_0^2}, \ \tilde{\mathbf q} = \frac{\mathbf q}{\rho_0 V_0^{3}}, \ \tilde{\mu} = \frac{\mu}{\rho_0 L_0 V_0}, \tilde{\mathbf v} = \frac{\mathbf v}{V_0}, \ \tilde f = \frac{f}{\rho_0 /V_0^3}, \ 
\end{aligned}
\end{equation}
where $\mu$ denotes the dynamic viscosity coefficient.
Physical quantities with a subscript 0 indicate their value in the reference state, where $V_0=\sqrt{2kT_0/m}$ is the most probable molecular speed.
The global Knudsen number is defined as
\begin{equation}
    \mathrm{Kn}=\frac{\ell_0}{L_0}=\frac{V_0}{L_0 \nu_0},
\end{equation}
where $\ell_0=V_0/\nu_0$ is the referenced molecular mean free path and $\nu_0$ is the mean collision frequency.
For brevity, we drop the tilde notation to denote dimensionless variables henceforth.



\subsection{Optimization of numerical flux}

Developing low-dissipation, strongly robust numerical fluxes is an important element of modern CFD.
The success of many numerical methods can be attributed to the combination of central and upwind discretizations, e.g., hybrid central-upwind schemes \cite{evje2005hybrid,kurganov2007reduction} and gas-kinetic schemes \cite{xu2014direct,sun2015explicit}.
The proportion of upwind and central contributions is usually adjustable to accommodate the mechanisms of convection and diffusion in different flows.
In addition to relying on a priori assumptions, the proportion can be determined by solving the optimization problem using $\partial P$.

Based on the definition in Eq.(\ref{eq:flux}), here we explicitly write the distribution function at the $k$-th face as 
\begin{equation}
    f_k^f = \chi f_{k}^u + (1-\chi) f_{k}^c,
\end{equation}
where $f_k^u$ and $f_k^c$ denote the particle distribution functions constructed by the upwind and central approaches, respectively.
The two distribution functions are unified by the coefficient $\chi \in [0,1]$.
We assume that the cell index to which the normal vector of face points is $i+1$ and the other cell corresponding to it is $i$, and the upwind distribution function can constructed according to Eq.(\ref{eq:flux}), i.e.,
\begin{equation}
\begin{aligned}
    &f_{k}^u = \hat f_{i}^k H(\mathbf n \cdot \mathbf v) + \hat f_{i+1}^k (1-H(\mathbf n \cdot \mathbf v)),
\end{aligned}
\label{eq:flux1}
\end{equation}
where $\{\hat f_{i}^k,\hat f_{i+1}^k\}$ are the reconstructed distribution functions on both sides of the face, and $H$ denotes the Heaviside step function.
The central contribution can be modeled as a Maxwellian distribution,
\begin{equation}
    f_{k}^c = \mathcal M_k^c,
\end{equation}
which can be determined with the help of the compatibility condition, i.e.,
\begin{equation}
    \int_{\mathbb R^3} \mathcal M_{k}^c \psi d\mathbf v = \int_{\mathbf n \cdot \mathbf v \ge 0} \hat f_i^k \psi d\mathbf v+ \int_{\mathbf n \cdot \mathbf v <0} \hat f_{i+1}^k \psi d\mathbf v.
\end{equation}
Note that incorporating a specific form of the distribution function in Eq.(\ref{eq:flux1}) leads to different gas dynamics.
Different truncation orders of the Chapman-Enskog expansion can yield the Euler, Navier-Stokes, and extended hydrodynamic solutions.

Here, we consider the construction of numerical fluxes for the Euler equations.
The particle distribution functions on both sides of the face adopt the Maxwellian determined by the reconstructed conservative variables,
\begin{equation}
    \hat f_i^k = \hat{\mathcal M}_i^k, \quad \hat f_{i+1}^k = \hat{\mathcal M}_{i+1}^k.
\end{equation}
The macroscopic fluxes for conservative variables are thus given by
\begin{equation}
    \mathbf F^W_k = \int_{\mathbb R^3} \mathbf v f_k^f\psi d\mathbf v.
    \label{eq:flux macro}
\end{equation}
Since the distribution function $f_k^f$ consists of three Maxwellian distributions, the above integral can be analytically solved.
The Sod shock tube problem is employed as the numerical experiment.
The initial particle distribution function is set as Maxwellian in correspondence with the following macroscopic variables,
\begin{equation}
\left(\begin{array}{c}
\rho \\
U \\
p \\
\end{array}\right)_{{t=0,x<0.5}}=\left(\begin{array}{c}
1 \\
0 \\
1 \\
\end{array}\right), \quad 
\left(\begin{array}{c}
\rho \\
U \\
p \\
\end{array}\right)_{{t=0,x\ge 0.5}}=\left(\begin{array}{c}
0.125 \\
0 \\
0.1 \\
\end{array}\right).
\end{equation}
The system is non-dimensionalized by the tube length together with the initial physical quantities on the left side.
The detailed computational setup is presented in Table \ref{tab:sod}, 
where $N_p$ indicates the number of trainable parameters.
\begin{table}[htbp]
    \caption{Computational setup of the Sod shock tube problem.} 
    \centering
    \begin{tabular}{lllllll} 
        \hline
        Equation & Gas & $t$ & $x$ & $N_{\mathbf x}$ & Order & Flux \\ 
        Euler & Argon & $(0,0.2]$ & $[0,1]$ & $200$ & 1 & Central-Upwind \\ 
        \hline
        Integrator & Boundary & CFL & $N_{p}$ & Optimizer \\
        Euler & Dirichlet & 0.5 & $\{1,199\}$ & AdamW \\
        \hline
    \end{tabular} 
    \label{tab:sod}
\end{table}

The flux function is optimized using two approaches. 
The first approach creates and optimizes a single parameter that controls the global behavior of the flux function in Eq.(\ref{eq:flux macro}).
The second strategy constructs a parameter at each face (199 independent parameters in total) that provides fine-grained control of local evolution.
To bound the predicted proportions of central and upwind contributions, the sigmoid function is employed to normalize the trainable parameters, i.e.,
\begin{equation}
    \chi = \sigma(\mathbf p).
\end{equation}
The initial value of $\mathbf p$ is set as $5.0$.
The cost function is defined as
\begin{equation}
    C = \sum_i^{N_x} \|\mathbf W_i^{\mathrm{ref}}(t=0.2)-\mathbf W_i(t=0.2)\|^2,
\end{equation}
where the reference solution $\mathbf W^\mathrm{ref}$ is obtained from the theoretical solution of the one-dimensional Riemann problem.

\begin{figure}[htb!]
    \centering
    \subfigure[Density]{
        \includegraphics[width=0.47\textwidth]{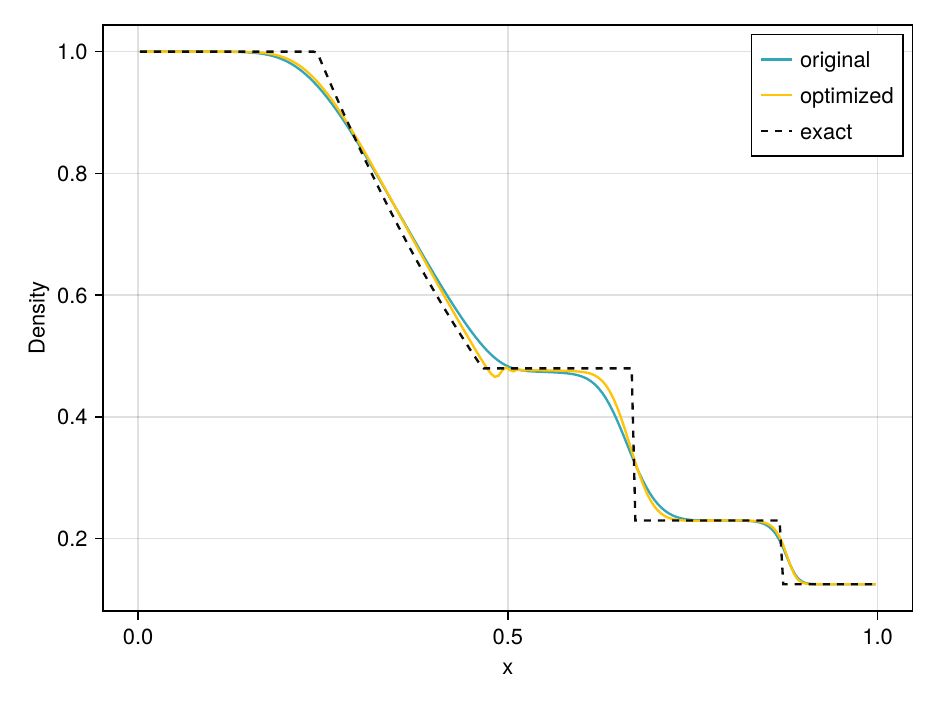}
    }
        \subfigure[Velocity]{
        \includegraphics[width=0.47\textwidth]{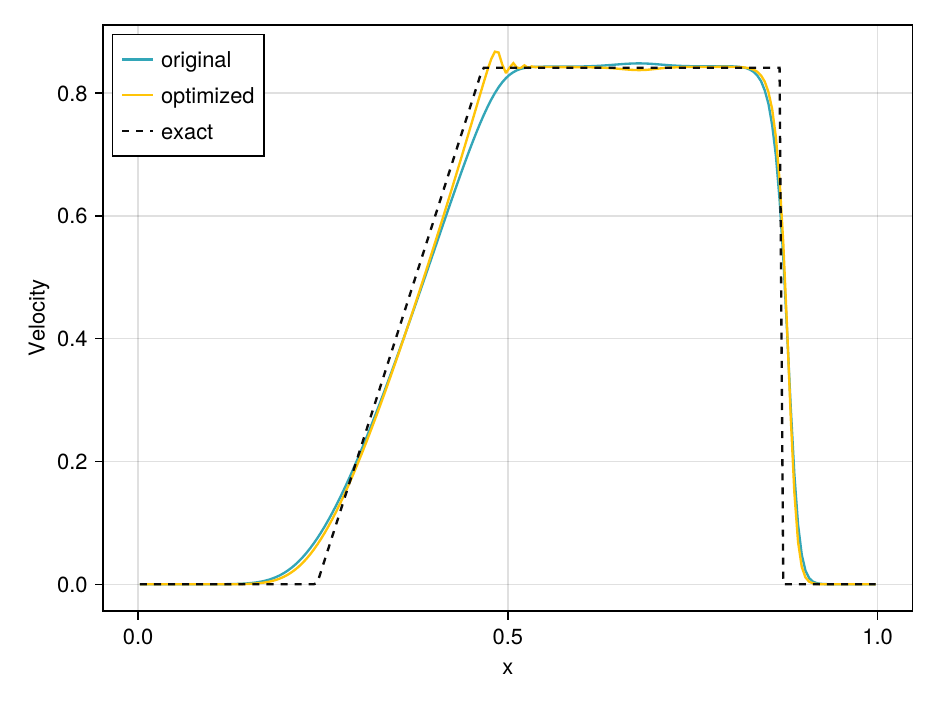}
    }
        \subfigure[Temperature]{
        \includegraphics[width=0.47\textwidth]{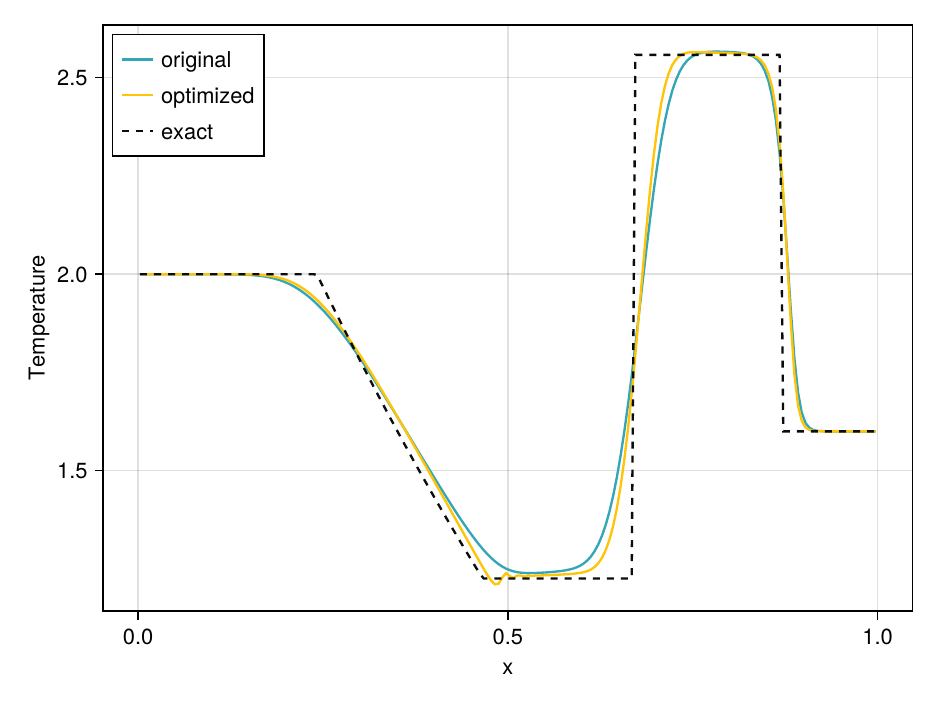}
    }
    \subfigure[Pressure]{
        \includegraphics[width=0.47\textwidth]{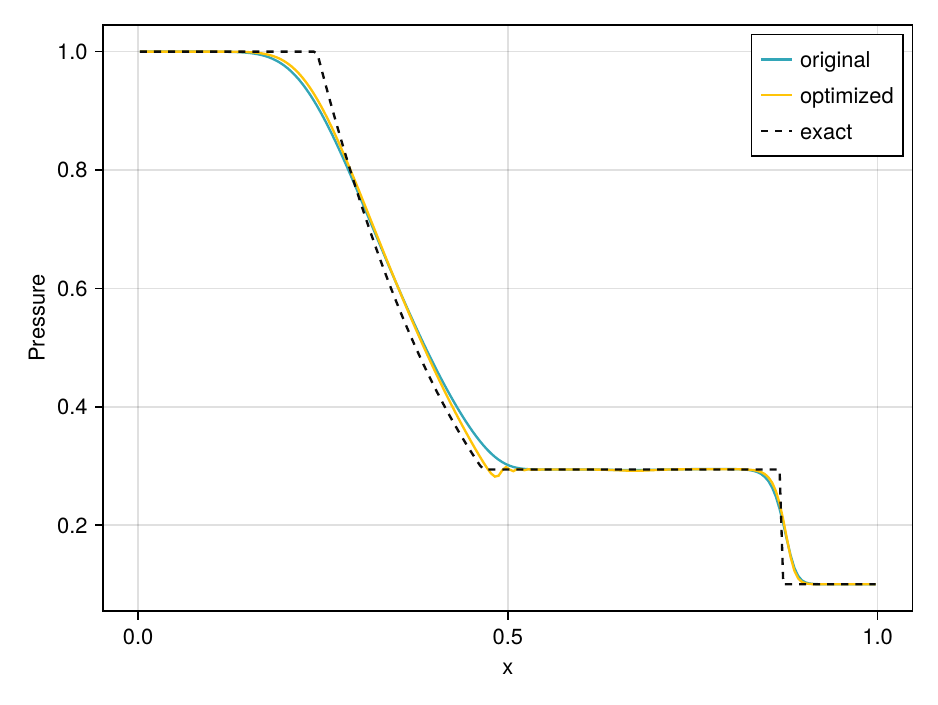}
    }
    \caption{Profiles of density, $U$-velocity, temperature, and pressure at $t=0.2$ simulated by the single-parameter model in the Sod shock tube problem.}
    \label{fig:sod single}
\end{figure}

\begin{figure}[htb!]
    \centering
    \subfigure[Density]{
        \includegraphics[width=0.47\textwidth]{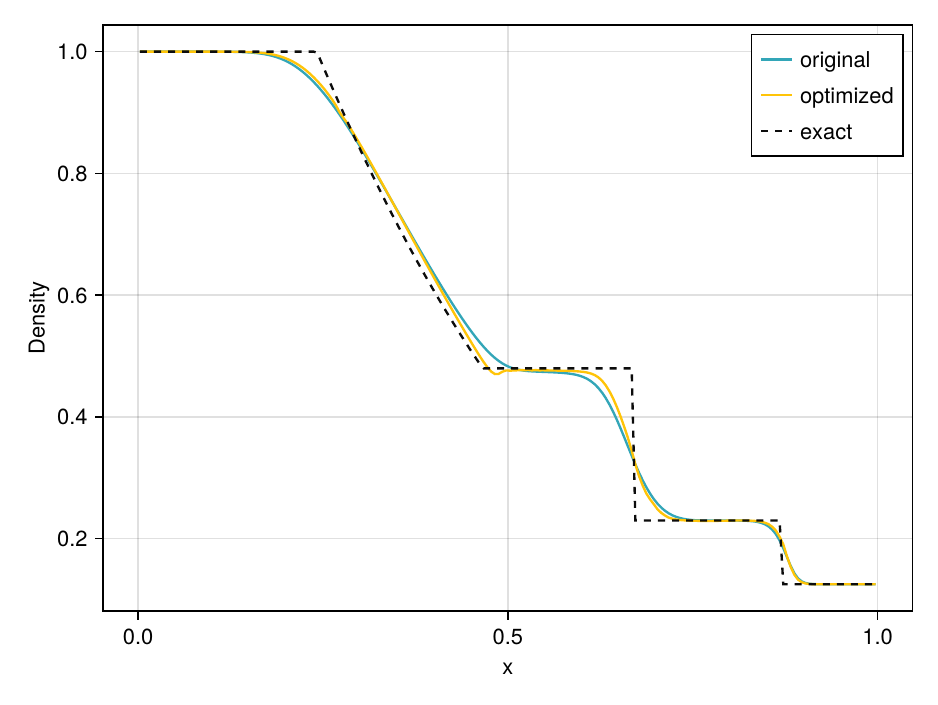}
    }
        \subfigure[Velocity]{
        \includegraphics[width=0.47\textwidth]{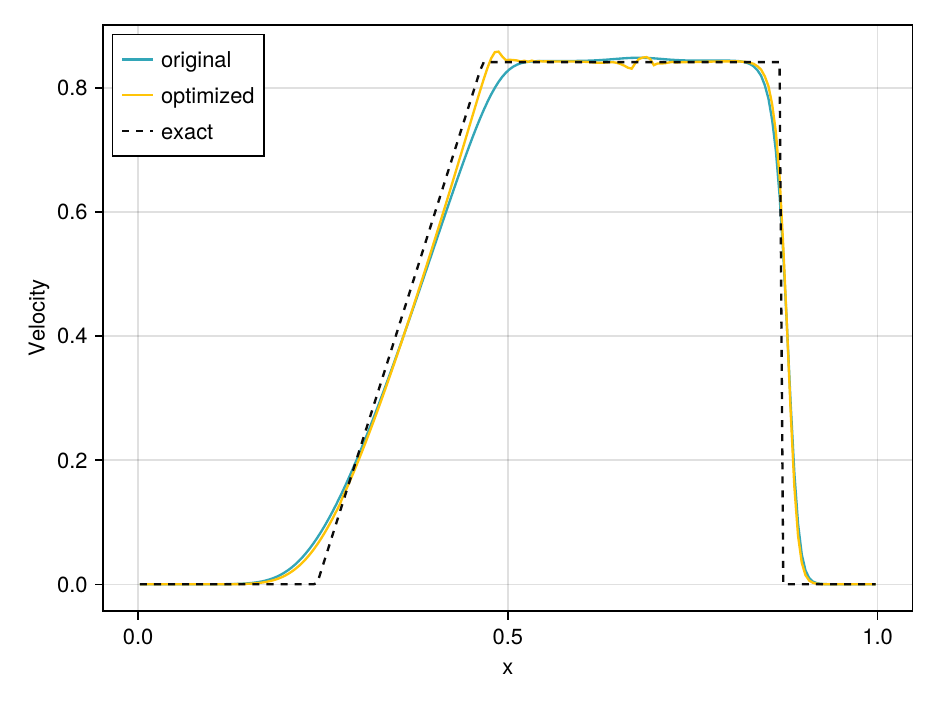}
    }
        \subfigure[Temperature]{
        \includegraphics[width=0.47\textwidth]{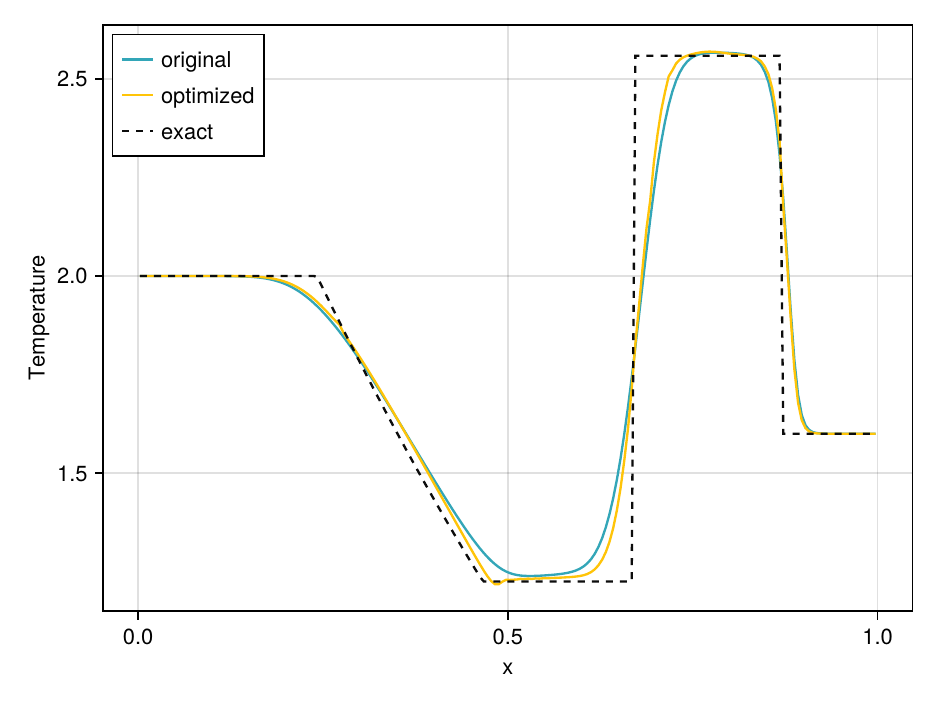}
    }
    \subfigure[Pressure]{
        \includegraphics[width=0.47\textwidth]{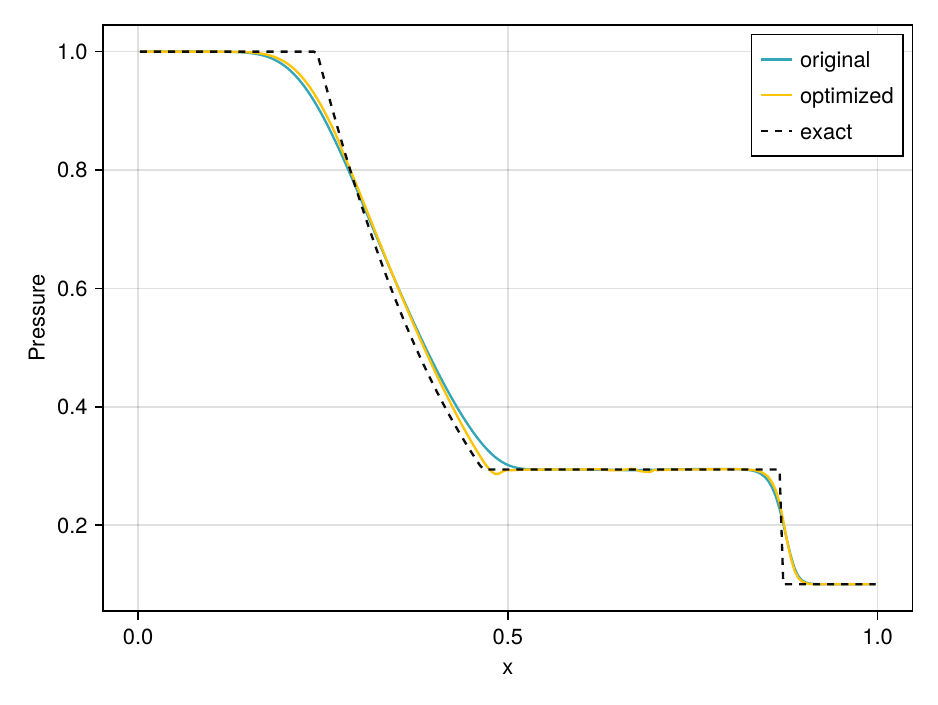}
    }
    \caption{Profiles of density, $U$-velocity, temperature, and pressure at $t=0.2$ simulated by the multi-parameter model in the Sod shock tube problem.}
    \label{fig:sod multi}
\end{figure}

\begin{figure}
    \centering
    \includegraphics[width=0.47\textwidth]{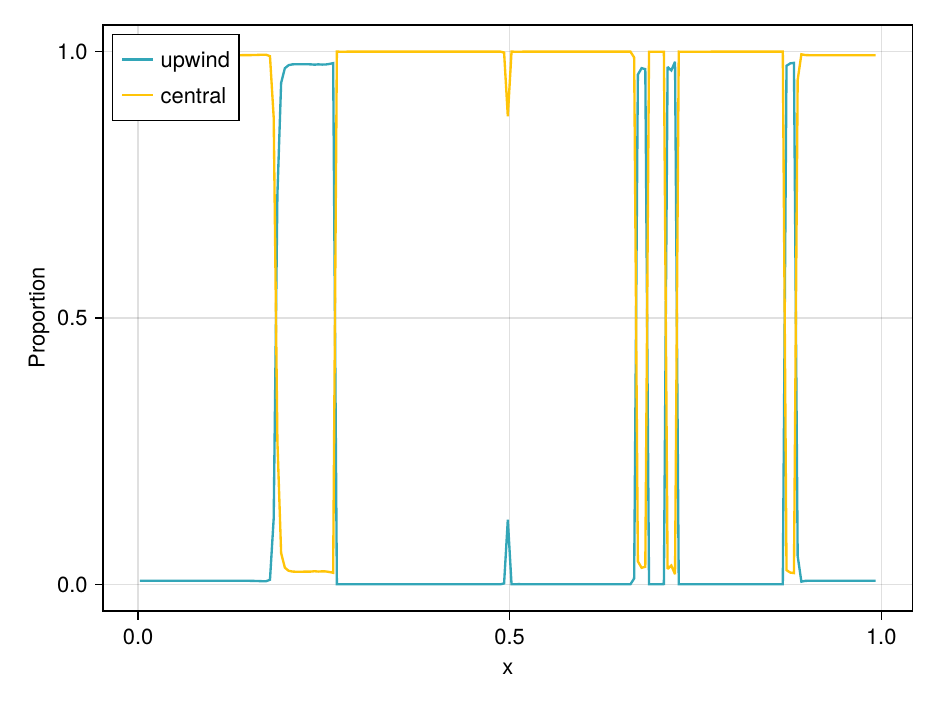}
    \caption{Proportions of contributions of the multi-parameter model in the Sod shock tube problem.}
    \label{fig:sod proportion}
\end{figure}

Figure \ref{fig:sod single} presents the profiles of density, velocity, temperature, and pressure in the shock tube at $t=0.2$ simulated by the single-parameter model.
Table \ref{tab:sod proportion} provide the contribution proportions of central and upwind fluxes before and after optimization.
It can be seen that as the dominant mechanism shifts from the upwind to the central scheme, the numerical dissipation in the numerical method is significantly reduced and the numerical solution is thus closer to the reference.
The optimized flux function indicates that less than $2.5\%$ of the upwind fluxes is sufficient to obtain and maintain robust discontinuous solutions.
Figure \ref{fig:sod multi} presents the numerical results of the multi-parameter model.
With the increased degrees of freedom due to multiple parameters, localized numerical dissipation can be better controlled.
As a result, the undershoot and oscillation near the tail of the rarefaction wave are mitigated, while the rest of the domain remains highly accurate.
Figure \ref{fig:sod proportion} shows the contribution proportions of central and upwind fluxes after optimization in the domain.
It is clear that the flux function increases the share of upwind contributions near the shock wave, contact discontinuity, and front region of the rarefaction wave, thus enhancing the robustness of the numerical scheme in these highly dissipative regions.
In other regions, the central-dominant flux effectively reduces the numerical dissipation of the scheme, ensuring that accurate physical solutions can be obtained.

\begin{table}[h]
    \centering
    \caption{Proportions of central and upwind fluxes of the single-parameter model in the Sod shock tube problem.}
    \begin{tabular}{lll}
        \toprule
        & Central & Upwind \\
        \midrule
        before optimization & 0.67\% & 99.33\% \\
        after optimization & $97.54\%$ & $2.46\%$ \\
        \bottomrule
    \end{tabular}
    \label{tab:sod proportion}
\end{table}




\subsection{Identification of fluid property}

Obtaining accurate fluid properties is a prerequisite for analysis and prediction.
Due to the limited measurement accuracy and the sparseness of sensors, physical parameters of gases often need to be obtained indirectly through inversion \cite{liu2019inverse}.
Among others, viscosity is an important property, which determines the strength of pressure and viscous effects in different flow regimes.
In this numerical experiment, a calibration problem of determining the viscosity coefficient from flow field data is considered.

Here we employ the hard-sphere model for monatomic gas.
The dynamic viscosity coefficient for hard-sphere gas can be determined as
\begin{equation}
    \mu = \mu_\mathrm{ref} \left(\frac{T}{T_\mathrm{ref}}\right)^\eta,
\end{equation}
where $\mu_\mathrm{ref}$ and $T_\mathrm{ref}$ denote the viscosity and temperature in the reference state, and $\eta=0.5$ is the viscosity index.
Once the viscosity is determined, the mean relaxation time can be obtained from the kinetic theory \cite{vincenti1966introduction}, i.e.,
\begin{equation}
    \tau = \frac{1}{\nu} =\frac{\mu}{p},
\end{equation}
which can then be used to solve the BGK model equation.



The wave propagation problem is employed as the numerical experiment.
The particle distribution function is initialized as Maxwellian, which corresponds to the following macroscopic variables,
\begin{equation}
\left(\begin{array}{c}
\rho \\
U \\
p \\
\end{array}\right)_{{t=0}}=\left(\begin{array}{c}
1+0.1\sin (2\pi x) \\
1 \\
0.5 \\
\end{array}\right).
\end{equation}
The system is non-dimensionalized by the domain length and initial unperturbed quantities.
The computational setup is listed in Table \ref{tab:wave}, where Tsit5 refers to Tsitouras’ 5/4 Runge-Kutta method \cite{tsitouras2011modified}.
To bound the prediction, the reference viscosity is set as the absolute value of the trainable parameter $\mathbf p \in\mathbb R^{N_p=1}$, i.e.,
\begin{equation}
    \mu_\mathrm{ref}=| \mathbf p |.
\end{equation}
The initial value of $\mathbf p$ is $10.0$.
The cost function is defined as
\begin{equation}
    C = \sum_i^{N_x}\sum_j^{N_v}\sum_k^{N_t} |f_{i,j}^{\mathrm{ref}}(t_k)-f_{i,j}(t_k)|^2,
\end{equation}
where the reference solution $f^\mathrm{ref}$ is the numerical solution at $\mu_\mathrm{ref}=0.01$.


\begin{table}[htbp]
    \caption{Computational setup of the wave propagation problem.} 
    \centering
    \begin{tabular}{lllllllll} 
        \hline
        Equation & Gas & $t$ & $x$ & $N_{x}$ & Order & $v$ & $N_{v}$ & $\eta$ \\ 
        BGK & Argon & $(0,0.25]$ & $[0,1]$ & $100$ & 1 & $[-5,5]$ & 48 & 0.5 \\ 
        \hline
        Flux & Quadrature & Integrator & Boundary & CFL & $N_{p}$ & Optimizer \\
        Upwind & Rectangular & Tsit5 & Periodic & 0.5 & $1$ & LBFGS \\
        \hline
    \end{tabular} 
    \label{tab:wave}
\end{table}

\begin{figure}[htb!]
    \centering
    \subfigure[Density]{
        \includegraphics[width=0.47\textwidth]{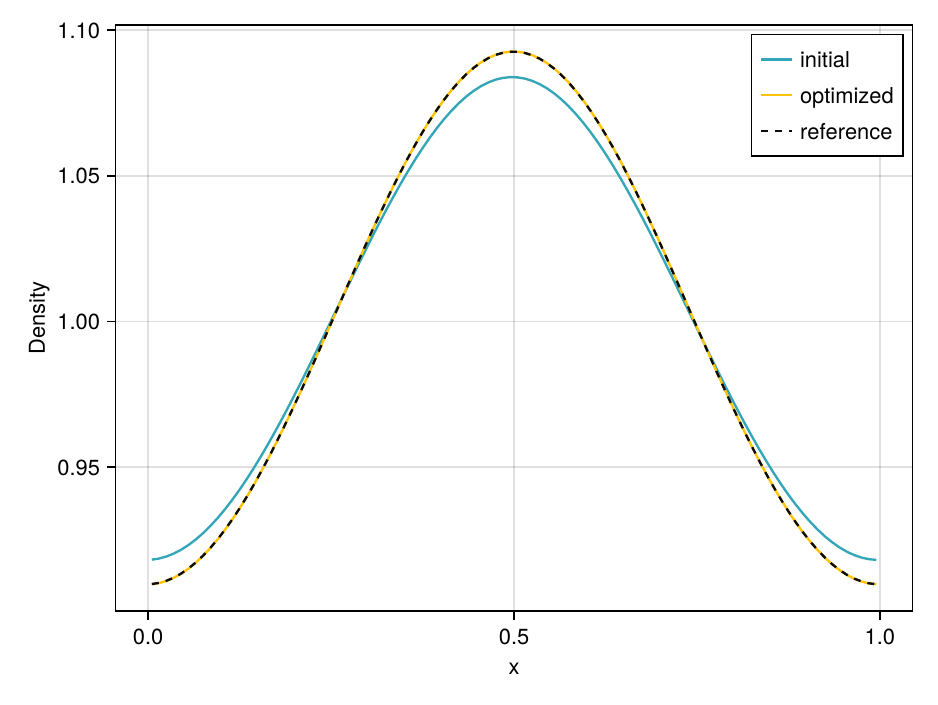}
    }
        \subfigure[Velocity]{
        \includegraphics[width=0.47\textwidth]{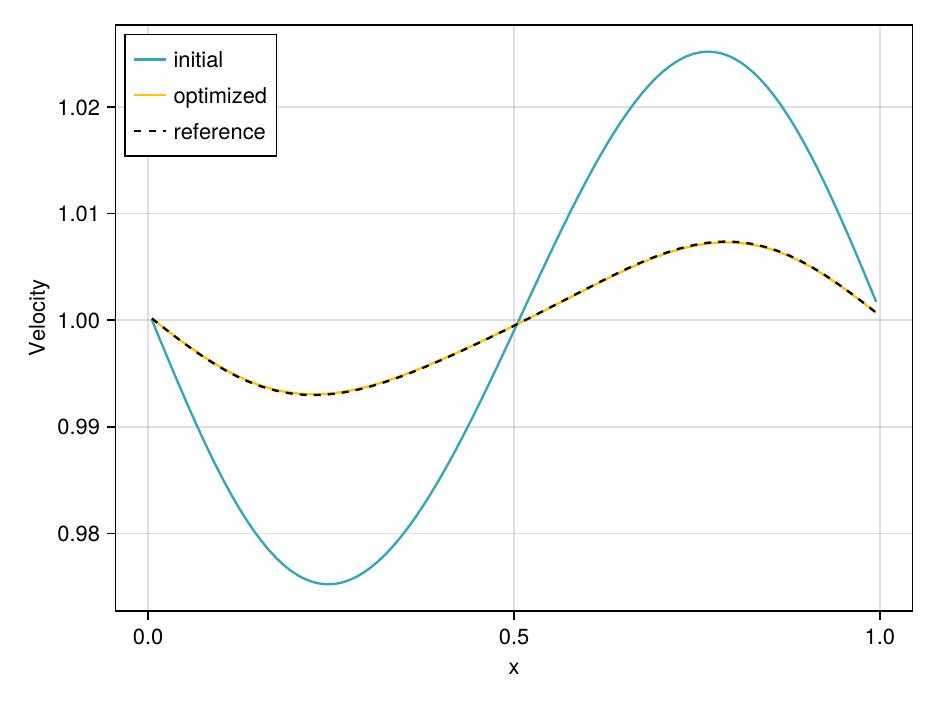}
    }
        \subfigure[Temperature]{
        \includegraphics[width=0.47\textwidth]{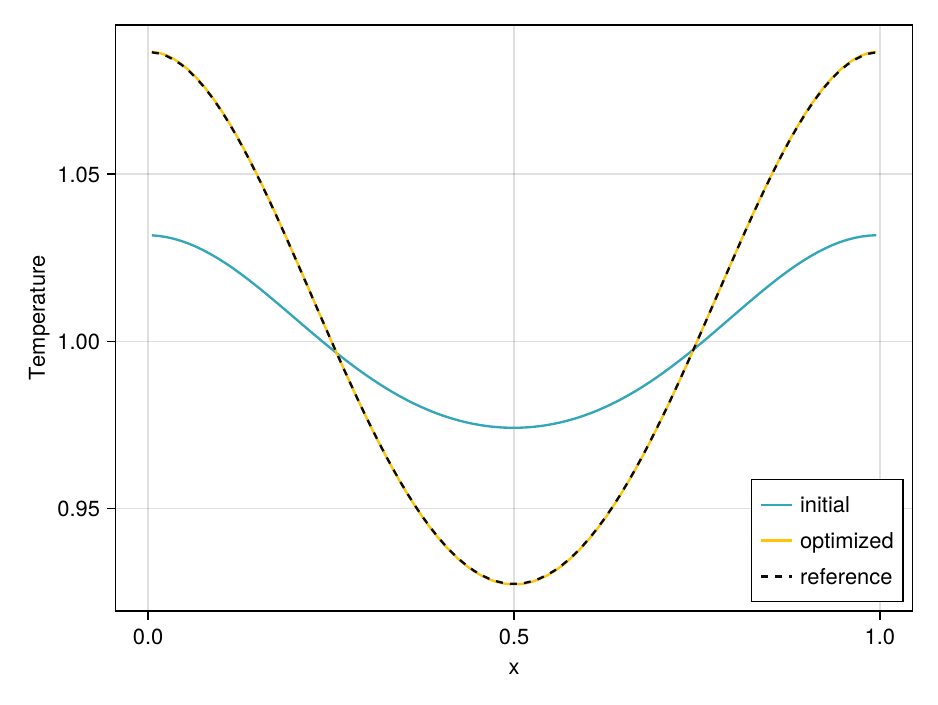}
    }
    \caption{Profiles of density, $U$-velocity, and temperature at $t=0.25$ simulated by the initial and optimized models in the wave propagation problem.}
    \label{fig:wave}
\end{figure}

Figure \ref{fig:wave} presents the density, velocity, and temperature profiles at $t=0.25$ simulated with the initial and optimized parameters.
The correct viscosity is recovered by aligning the trajectories of the numerical solution and reference target.
Table \ref{tab:wave value} shows the solution of the parameter in the optimization problem.
It can be seen that the accuracy of the solution is more than 98\%.
To illustrate the superiority of $\partial P$-based solution algorithm, we compare its performance with the ensemble Kalman inversion (EKI) \cite{iglesias2013ensemble,kovachki2019ensemble}.
It leverages the principles of the ensemble Kalman filter within the framework of the Bayesian inverse problem and is one of the state-of-the-art gradient-free methods for solving optimization problems.
To accelerate the convergence of the optimization problem, we incorporate the prior normal distribution $\mathbf p \sim \mathcal N(0,0.1^2)$ to sample the initial ensemble of parameters.
Table \ref{tab:wave cost} provides the computational costs until the cost value reduces to $C=0.00001$ based on $\partial P$- and EKI, respectively.
It can be seen that even in the case of strong intervention (by manually presetting parameter distributions closer to the true value), the computational time and allocations of EKI are still more than an order of magnitude higher than that of $\partial P$.
This indicates the effectiveness and necessity of developing $\partial P$-based solution algorithms.


\begin{table}[h]
    \centering
    \caption{Initial, optimized, and target values of dynamic viscosity coefficient in the wave propagation problem.}
    \begin{tabular}{lll}
        \toprule
        Target & Optimized & Initial \\
        \midrule
        0.01 & 0.00986 & 10 \\
        \bottomrule
    \end{tabular}
    \label{tab:wave value}
\end{table}

\begin{table}[h]
    \centering
    \caption{Computational costs of $\partial P$-based solution algorithm and ensemble Kalman inversion in the wave propagation problem.}
    \begin{tabular}{lll}
        \toprule
        & Time (s) & Allocation (GB) \\
        \midrule
        EKI & 455.61 & 434.54 \\
        $\partial P$ & 26.26 & 10.08 \\
        \bottomrule
    \end{tabular}
    \label{tab:wave cost}
\end{table}

\subsection{Construction of hydrodynamic closure}

Due to the high dimensionality and strong nonlinearity of the Boltzmann equation, numerous efforts have been devoted to extending the applicability of hydrodynamic models in non-equilibrium flow regimes.
The core task here is to construct reliable algebraic or evolutionary models for higher-order moment variables to approximate the particle distribution function, through which Eq.(\ref{eq:conservation laws}) becomes solvable.
It is challenging since the particle distribution information has been partially filtered out during the coarse-grained upscaling modeling processes.
Established theoretical work includes the Burnett and Super-Burnett equations based on the asymptotic Chapman-Enskog expansion, as well as moment equations based on monomials and polynomials hierarchies.
Due to the aforementioned challenge, these efforts have achieved limited success within specific flow regimes.

Neural networks, as a multi-parameter model, provide an alternative for constructing hydrodynamic closures through a data-driven approach.
Here, we construct a neural network-based constitutive model based on the Navier-Stokes equations that can depict non-equilibrium flows more accurately.
In the case of monatomic gas, for example, the constitutive relations in the Navier-Stokes equations include the generalized Newton's law and Fourier's law, i.e.,
%
\begin{equation}
\begin{aligned}
    &\mathbf P=-p \mathbf I + \mathbf T, \\
    &\mathbf T=2\mu (\nabla_{\mathbf x} \mathbf V+(\nabla_{\mathbf x} \mathbf V)^T)-\frac{2}{3}\mu (\nabla_{\mathbf x} \cdot \mathbf V) \mathbf I, \\
    &\mathbf q = -\kappa \nabla_{\mathbf x} T,
\end{aligned}
\end{equation}
where $\mathbf T$ is the stress tensor and $\mathbf I$ refers to the unit tensor.
In the computational framework of the finite volume method, the constitutive relations are usually incorporated in the flux function for ease of computation, e.g., the flux splitting scheme \cite{wada1997accurate} and gas-kinetic flux solver \cite{xu2001gas}.
Thus, we organize the mechanical-neural model through the flux function, i.e.,
\begin{equation}
    \mathbf F^W=\mathbf F^\mathrm{NS}+\mathbf F^\mathrm{NN},
\end{equation}
where $\mathbf F^\mathrm{NS}$ denotes the Navier-Stokes fluxes simulated by the gas-kinetic scheme, and $\mathbf F^\mathrm{NN}$ refers to their deviation from ground-truth non-equilibrium flow physics.
The output values of the neural network are equal to $\mathbf F^\mathrm{NN}$, while its inputs include macroscopic variables $\mathbf W$, their gradients $\nabla_{\mathbf x} \mathbf W$, and the Knudsen number $\mathrm{Kn}$ in the reference state.







The shear layer problem is employed as the numerical experiment. 
The flow field is initialized as
\begin{equation}
    \left(\begin{array}{c}
    \rho \\
    U \\
    V \\
    T \\
    \end{array}\right)_{{t=0,x<0}}=\left(\begin{array}{c}
    1 \\
    0 \\
    1 \\
    1 \\
    \end{array}\right), \quad 
    \left(\begin{array}{c}
    \rho \\
    U \\
    V \\
    T \\
    \end{array}\right)_{{t=0,x\ge 0}}=\left(\begin{array}{c}
    1 \\
    0 \\
    -1 \\
    0.5 \\
    \end{array}\right).
\end{equation}
The initial particle distribution function is Maxwellian everywhere in correspondence to the macroscopic variables.
The system is non-dimensionalized by the domain length and physical quantities on the left side.
The computational setup is provided in Table \ref{tab:layer},
where $\tau_0$ denotes the mean relaxation time in the reference state.
A fully connected neural network with 420 parameters is employed to build the modified flux function.
The cost function is defined as
\begin{equation}
    C = \sum_i^{N_x}\sum_j^{N_t} \|\mathbf W_{i}^{\mathrm{ref}}(t_j)-\mathbf W_{i}(t_j)\|^2 + \epsilon \|\boldsymbol{\theta}\|^2,
\end{equation}
where the reference solution $\mathbf W^\mathrm{ref}$ is obtained by solving the BGK kinetic equation and applying moments to the distribution function.

\begin{table}[htbp]
    \caption{Computational setup of the shear layer problem.} 
    \centering
    \begin{tabular}{lllllll} 
        \hline
        Equation & Gas & $t$ & $x$ & $N_x$ & Order & Flux \\ 
        Extended NS & Argon & $(0,10\tau_0]$ & $[-0.1,0.1]$ & $100$ & 1 & GKS+NN \\ 
        \hline
        Integrator & Boundary & Kn & CFL & $N_p$ & $\epsilon$ & Optimizer \\
        Euler & Dirichlet & 0.005 & 0.5 & $420$ & $10^{-5}$ & AdamW \\
        \hline
    \end{tabular} 
    \label{tab:layer}
\end{table}

\begin{figure}[htb!]
    \centering
    \subfigure[Density]{
        \includegraphics[width=0.47\textwidth]{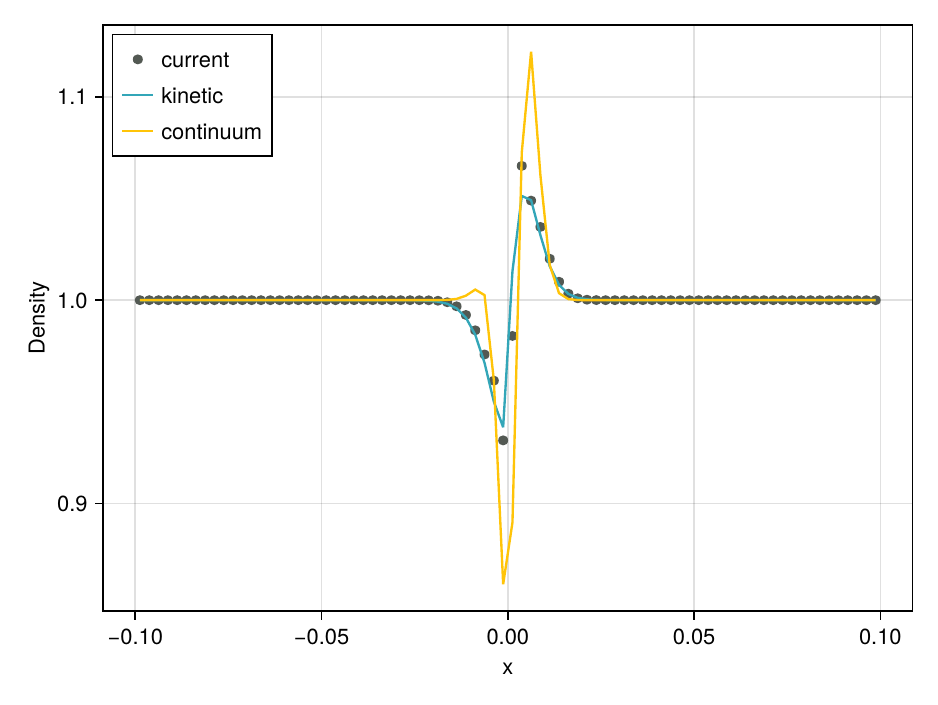}
    }
        \subfigure[$U$-velocity]{
        \includegraphics[width=0.47\textwidth]{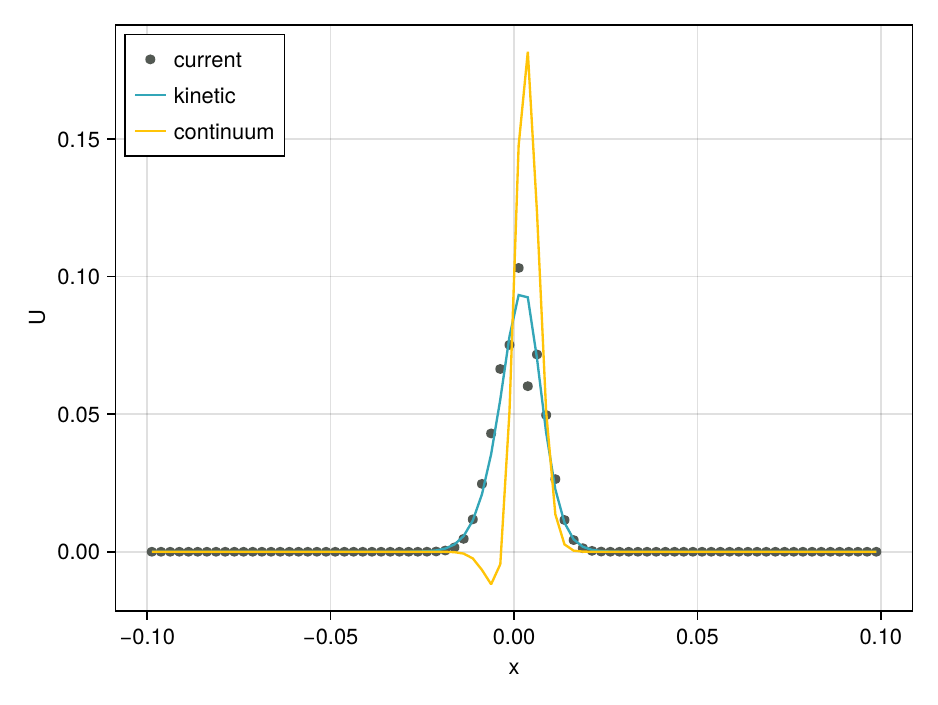}
    }
        \subfigure[$V$-velocity]{
        \includegraphics[width=0.47\textwidth]{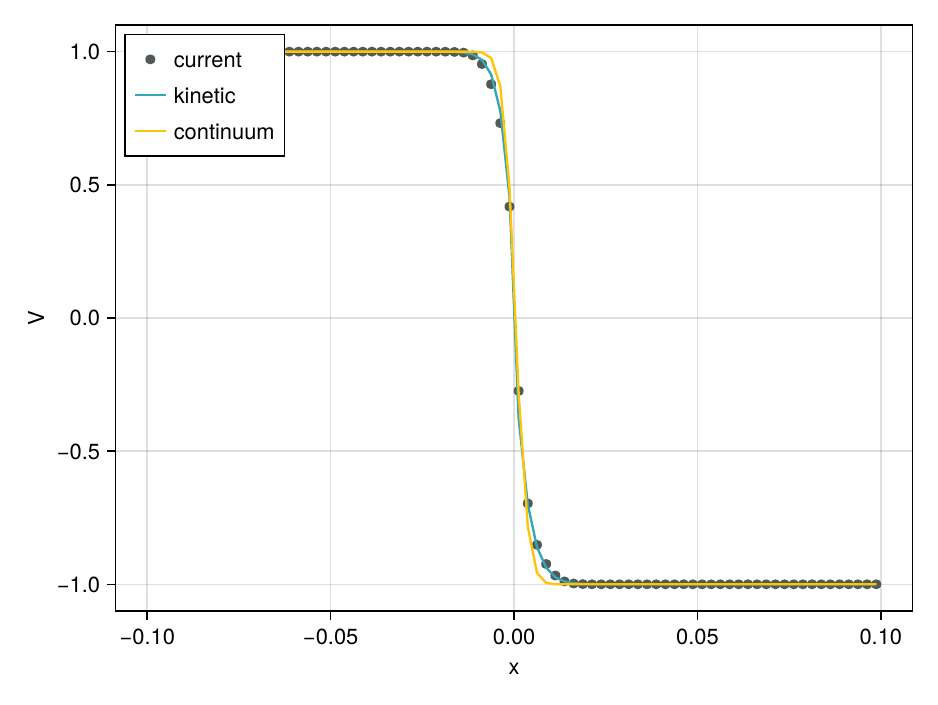}
    }
        \subfigure[Temperature]{
        \includegraphics[width=0.47\textwidth]{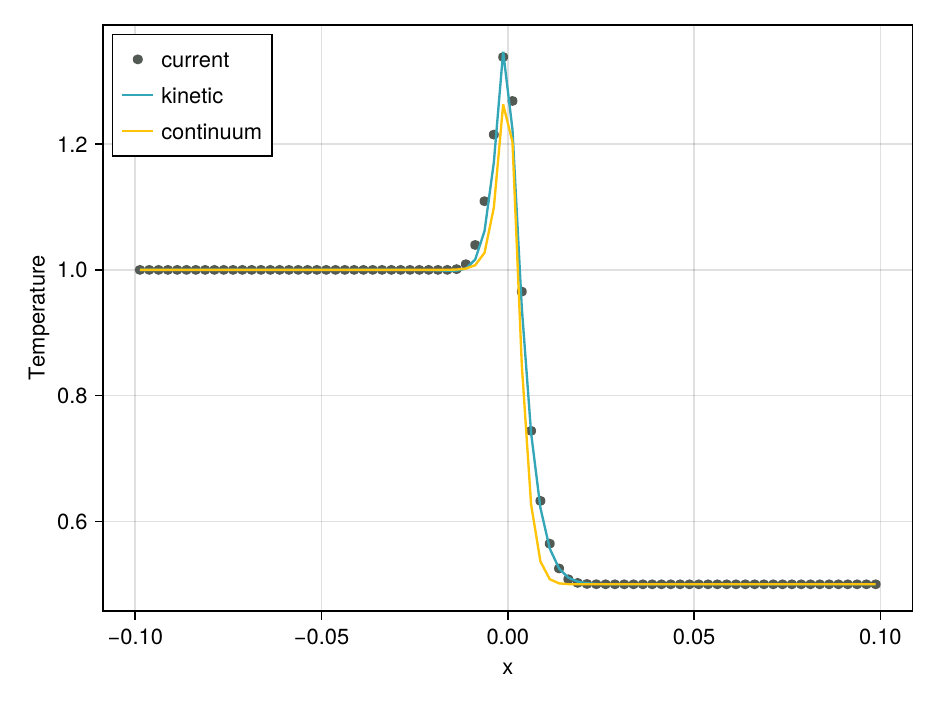}
    }
    \caption{Profiles of density, $U$-velocity, $V$-velocity, and temperature at $t=\tau_0$ simulated by different models in the shear layer problem.}
    \label{fig:layer t1}
\end{figure}

\begin{figure}[htb!]
    \centering
    \subfigure[Density]{
        \includegraphics[width=0.47\textwidth]{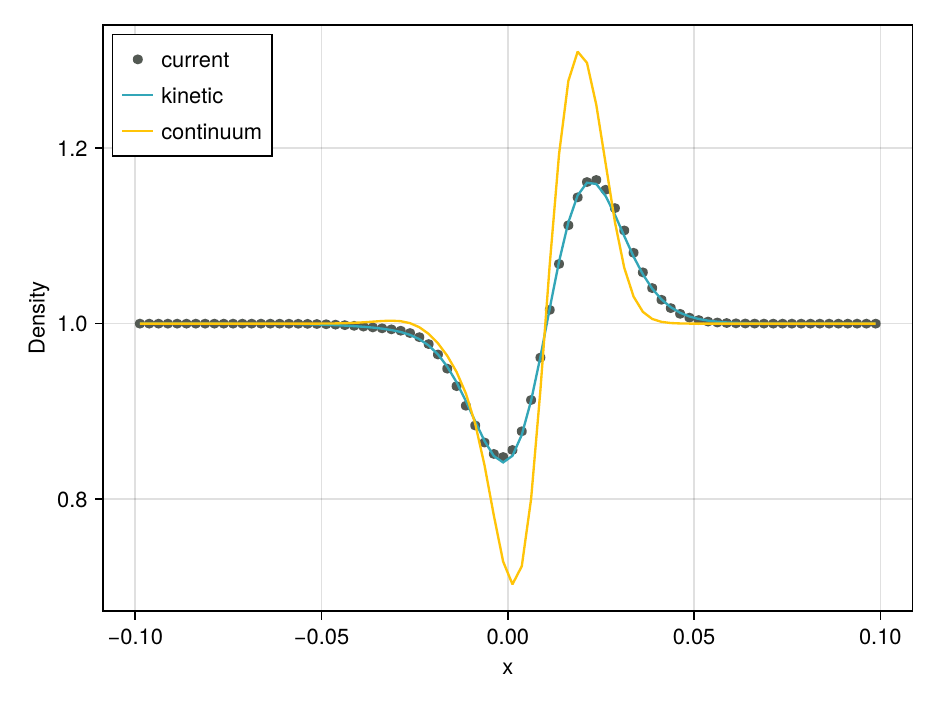}
    }
        \subfigure[$U$-velocity]{
        \includegraphics[width=0.47\textwidth]{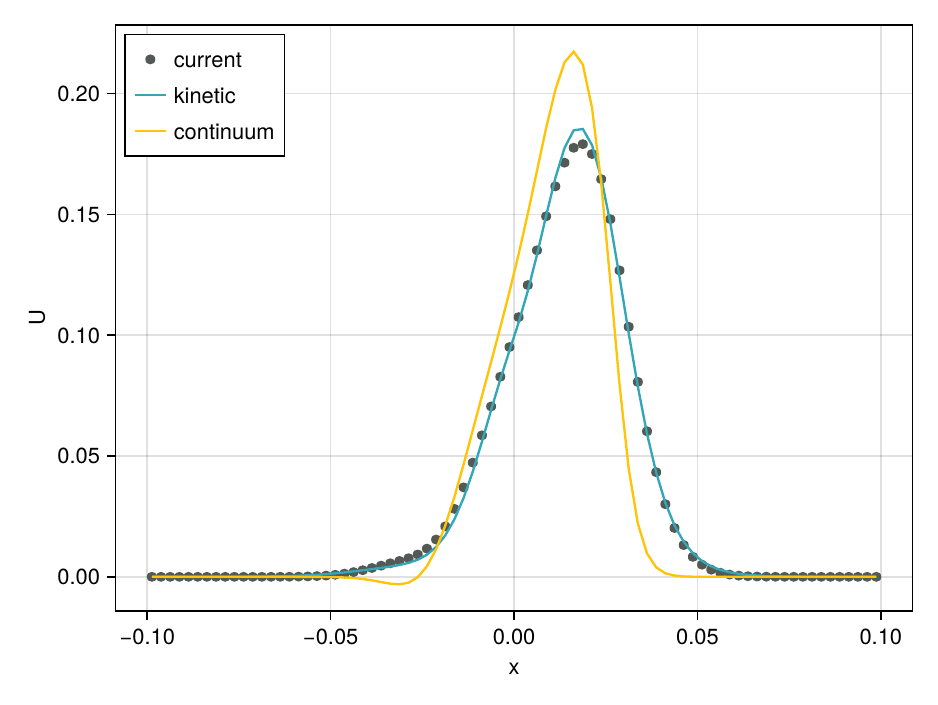}
    }
        \subfigure[$V$-velocity]{
        \includegraphics[width=0.47\textwidth]{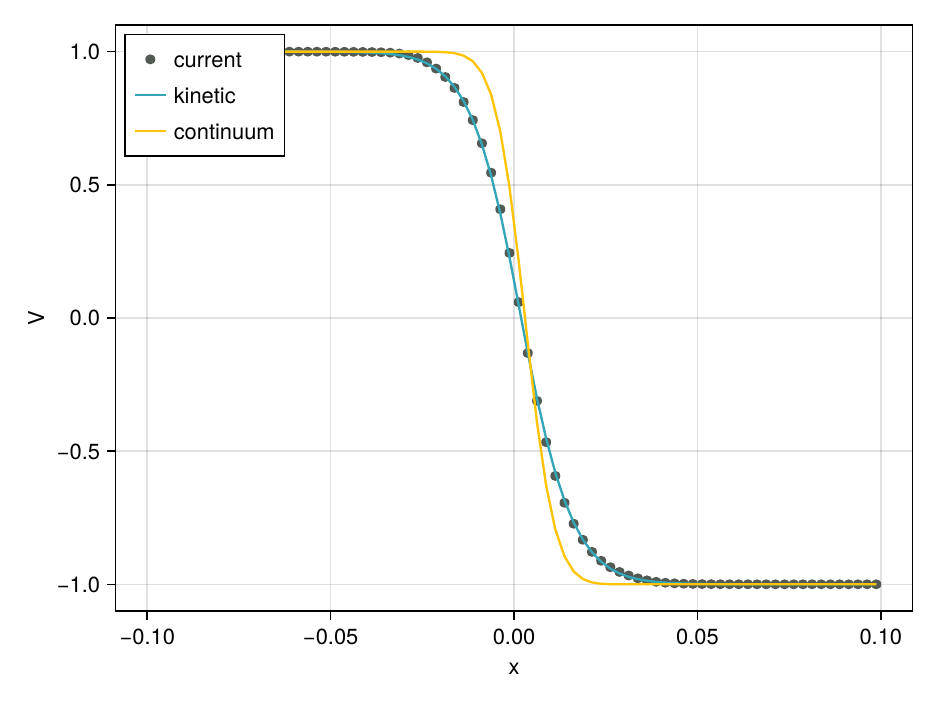}
    }
        \subfigure[Temperature]{
        \includegraphics[width=0.47\textwidth]{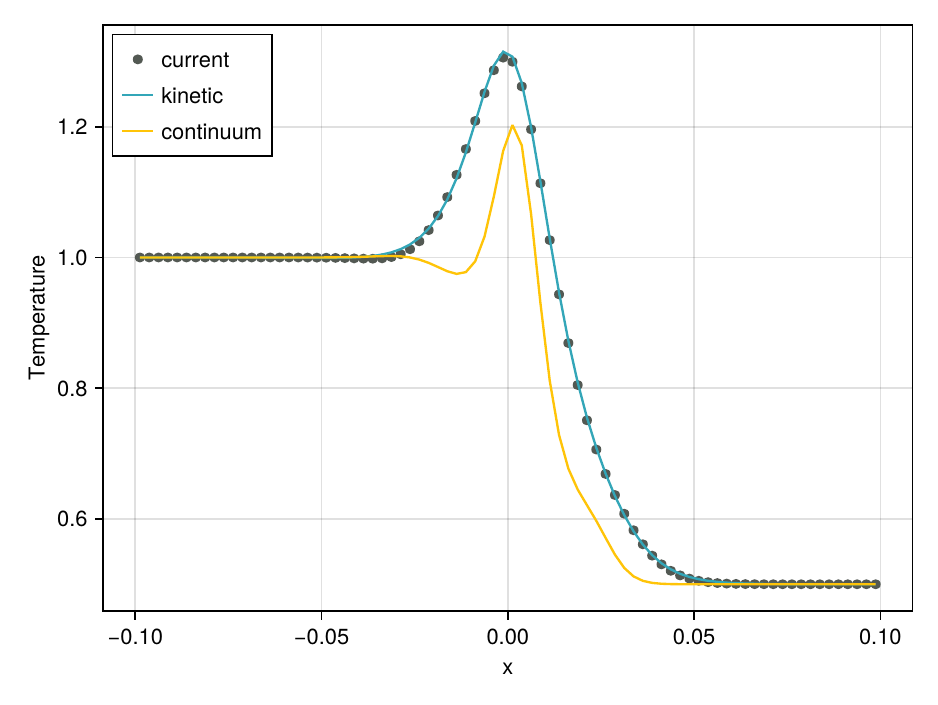}
    }
    \caption{Profiles of density, $U$-velocity, $V$-velocity, and temperature at $t=5\tau_0$ simulated by different models in the shear layer problem.}
    \label{fig:layer t2}
\end{figure}

\begin{figure}[htb!]
    \centering
    \subfigure[Density]{
        \includegraphics[width=0.47\textwidth]{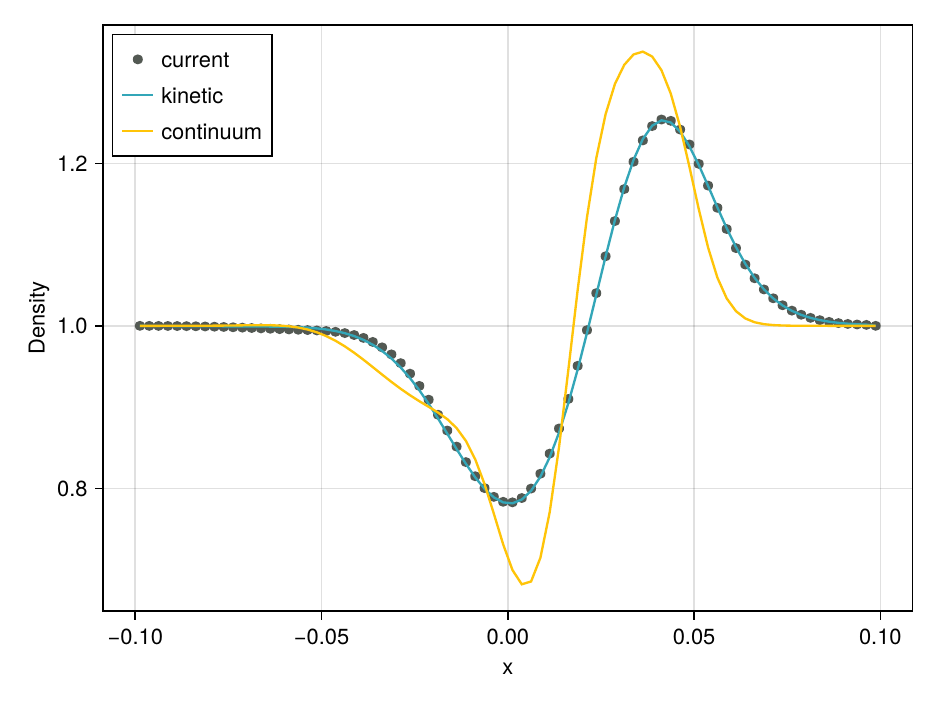}
    }
        \subfigure[$U$-velocity]{
        \includegraphics[width=0.47\textwidth]{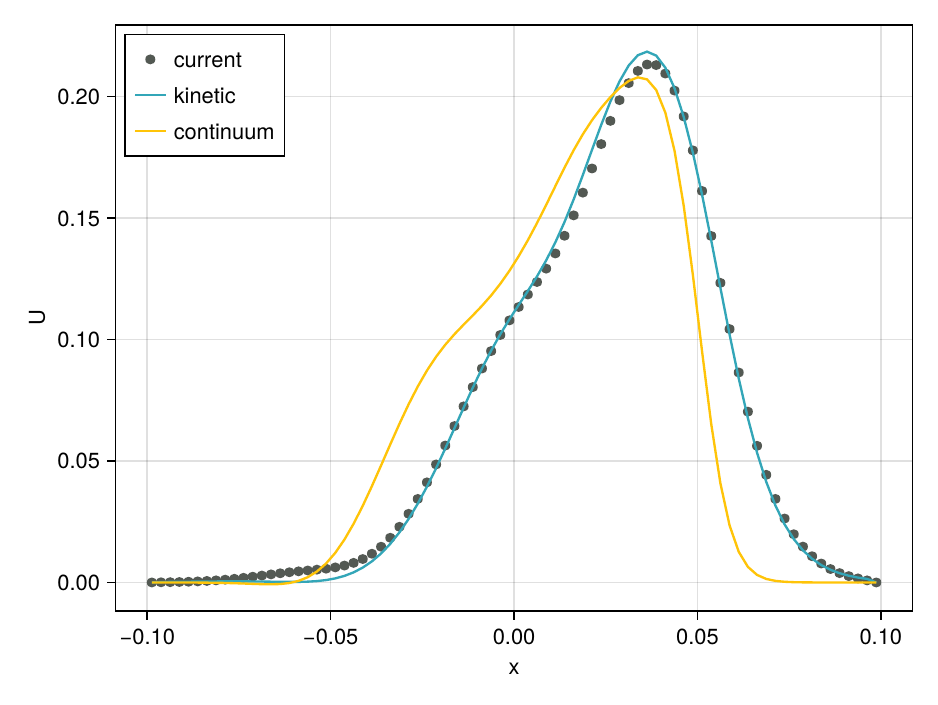}
    }
        \subfigure[$V$-velocity]{
        \includegraphics[width=0.47\textwidth]{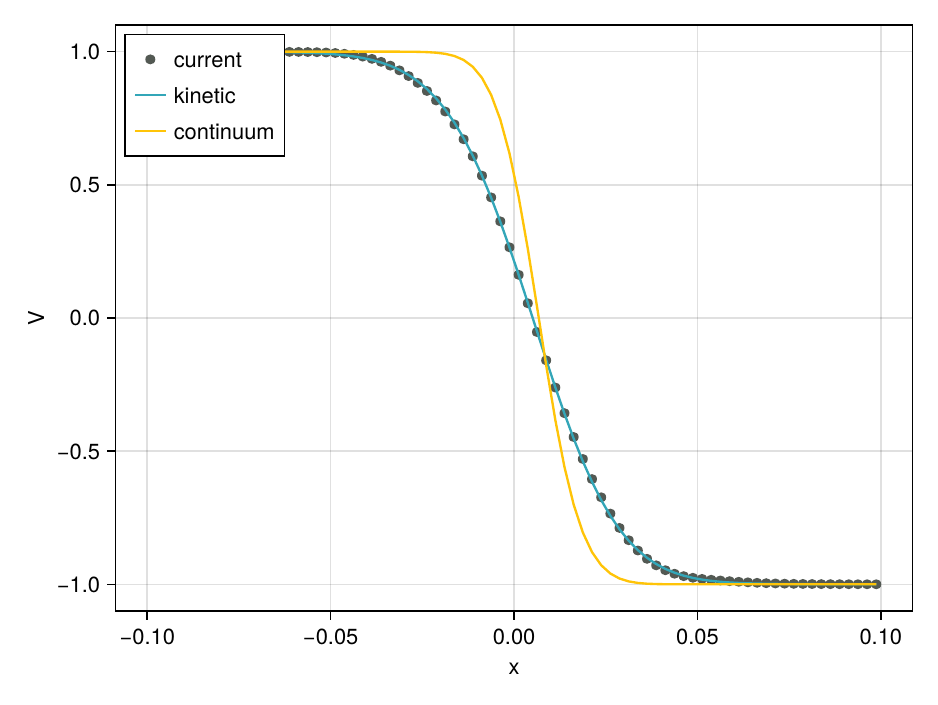}
    }
        \subfigure[Temperature]{
        \includegraphics[width=0.47\textwidth]{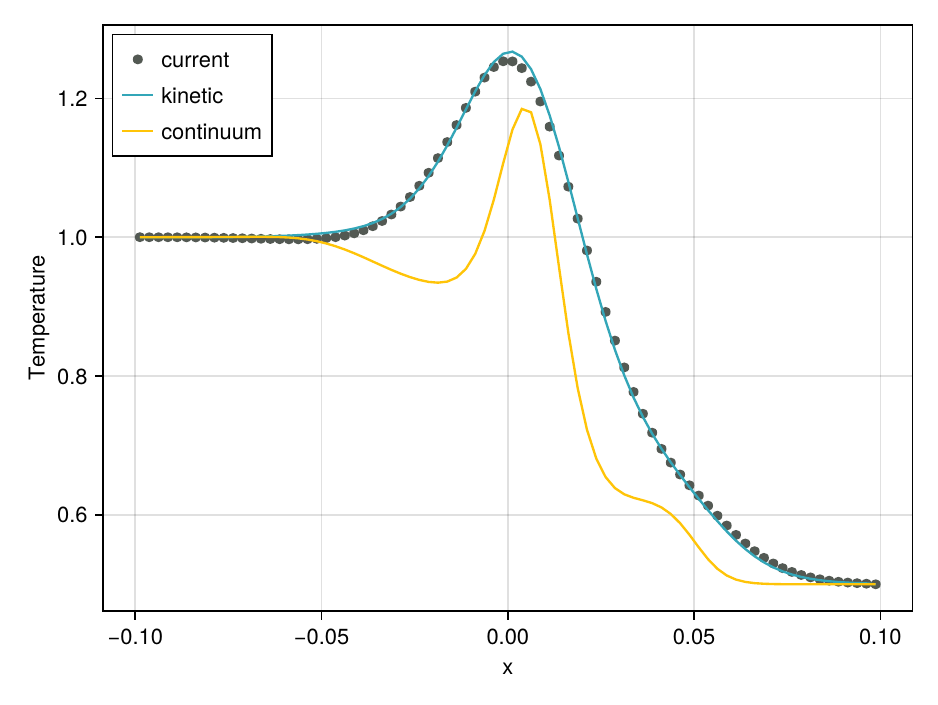}
    }
    \caption{Profiles of density, $U$-velocity, $V$-velocity, and temperature at $t=10\tau_0$ simulated by different models in the shear layer problem.}
    \label{fig:layer t3}
\end{figure}

Figure \ref{fig:layer t1}, \ref{fig:layer t2} and \ref{fig:layer t3} present the profiles of macroscopic flow variables at $t=\tau_0$, $5\tau_0$, and 10$\tau_0$.
The pressure-driven transport of momentum and energy forms a transition layer that thickens as time evolves.
As the results show, due to the lack of effective non-equilibrium constitutive relations, the Navier-Stokes equations predict a narrower transition layer along with greater density fluctuations.
With the supplement of the neural network-based closure model, non-equilibrium effects are well described within the framework of hydrodynamic equations, and the rate and pattern of viscous transport, which are identical to those of the BGK equation, are accurately recovered.
Table \ref{tab:layer cost} presents the computational costs of a single computation of Navier-Stokes fluxes and neural network inference in the $\partial P$-based solution algorithm.
It can be seen that the mechanical-neural model dramatically improves the accuracy of hydrodynamic equations while adding only around 25\% additional computational overhead.

\begin{table}[h]
    \centering
    \caption{Computational costs of Navier-Stokes fluxes and neural network model in the $\partial P$-based solution algorithm in the shear layer problem.}
    \begin{tabular}{lll}
        \toprule
        & Time ($10^{-4}$ s) & Allocation (KB) \\
        \midrule
        Navier-Stokes & 1.43 & 13.36 \\
        Neural Network & 0.37 & 5.20 \\
        \bottomrule
    \end{tabular}
    \label{tab:layer cost}
\end{table}

\subsection{Operator learning for the kinetic equation}

\begin{figure}
    \centering
    \includegraphics[width=0.6\textwidth]{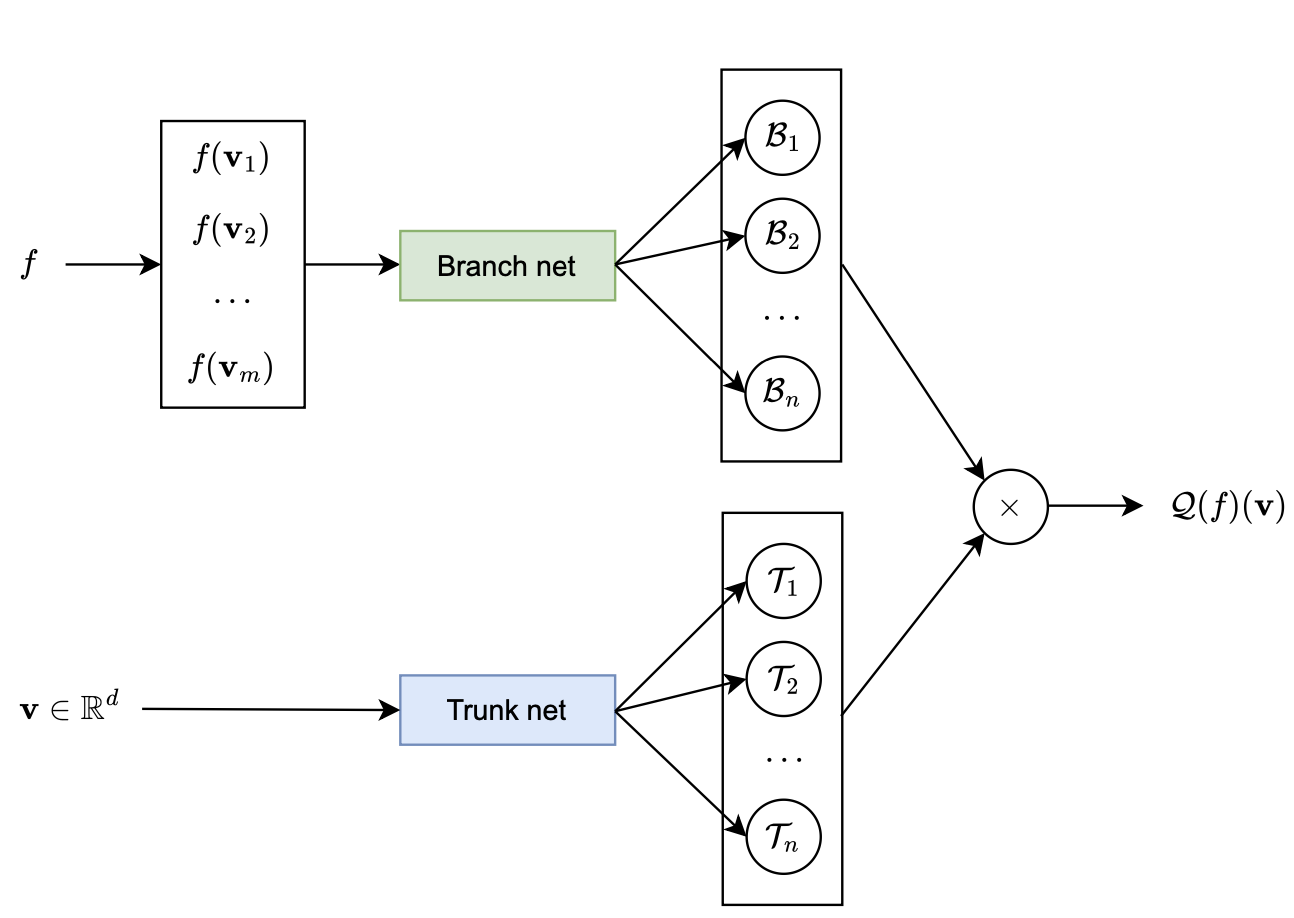}
    \caption{Architecture of DeepONet surrogate model of the Boltzmann equation.}
    \label{fig:deeponet}
\end{figure}

Due to the ability of neural networks in feature identification and dimension reduction, an alternative to solving non-equilibrium flows is to directly solve the Boltzmann equation with the help of neural networks.
Since the complexity of the algorithm for solving the Boltzmann equation lies mainly in the fivefold collision integral (larger than $O(N_v^6)$ for naive point-to-point computation), we construct the surrogate model for this operator based on the neural network.
The model employed here is the deep operator network (DeepONet) \cite{lu2021learning}, which is a neural architecture designed to learn nonlinear operator mappings between function spaces of infinite dimension.
The model consists of two subnetworks, i.e., a branch net $\mathcal B$, which encodes the input functions evaluated at fixed sensor points into a finite-dimensional representation in the latent space, and a trunk net $\mathcal T$, which encodes the locations at which the output function is evaluated.
The outputs of these networks are combined, typically via a dot product, to yield the output function's value at the specified location.
In this case, the input function to $\mathcal B$ is set as the particle distribution function evaluated at $N_v$ collocation points in the velocity space, i.e., $\{f_j\}_{j=1:N_v}$, and the desired output is the collision term of the Boltzmann equation evaluated at the input velocity point $\mathbf v\in \mathbb R^d$, where $d$ is the flow dimension of interest.
The architecture of the DeepONet model is shown in Figure \ref{fig:deeponet}.
The DeepONet-enhanced Boltzmann equation can be written as
\begin{equation}
    \frac{\partial f}{\partial t} + \mathbf v \cdot \nabla_\mathbf x f = \frac{1}{\mathrm{Kn}} \mathrm{NN}_{\boldsymbol{\theta}}(f).
\label{eq:ube}
\end{equation}


\subsubsection{Relaxation of non-equilibrium distribution}

We first consider the relaxation of a non-equilibrium distributed many-particle system.
The initial particle distribution function is set as
\begin{equation}
    f = \frac{1}{2}\left(\frac{1}{\pi}\right)^{3/2}(\exp(-(u-1)^2)+0.7\exp(-(u+1)^2))\exp(-v^2)\exp(-w^2).
\end{equation}
We are mainly concerned with non-equilibrium effects in the $x$-direction, which can be characterized as
\begin{equation}
    h=\int_{-\infty}^\infty \int_{-\infty}^\infty f dvdw.
\end{equation}
Based on the DeepONet, we expect to construct a one-dimensional model for the reduced distribution function $h$ that can recover correct three-dimensional effects.
The computational setup is detailed in Table \ref{tab:relaxation},
The DeepONet model employed consists of two fully connected neural networks with a total of 51600 parameters.
The cost function is defined as
\begin{equation}
    C = \sum_j^{N_v}\sum_k^{N_t} |f_{j}^{\mathrm{ref}}(t_k)-f_{j}(t_k)|^2,
\end{equation}
where the reference solution $f^\mathrm{ref}$ is obtained by solving the Boltzmann equation with the fast spectral method \cite{mouhot2006fast,wu2013deterministic} and then projecting it to one-dimensional space.

\begin{figure}[htb!]
    \centering
    \subfigure[$t=0.1$]{
        \includegraphics[width=0.47\textwidth]{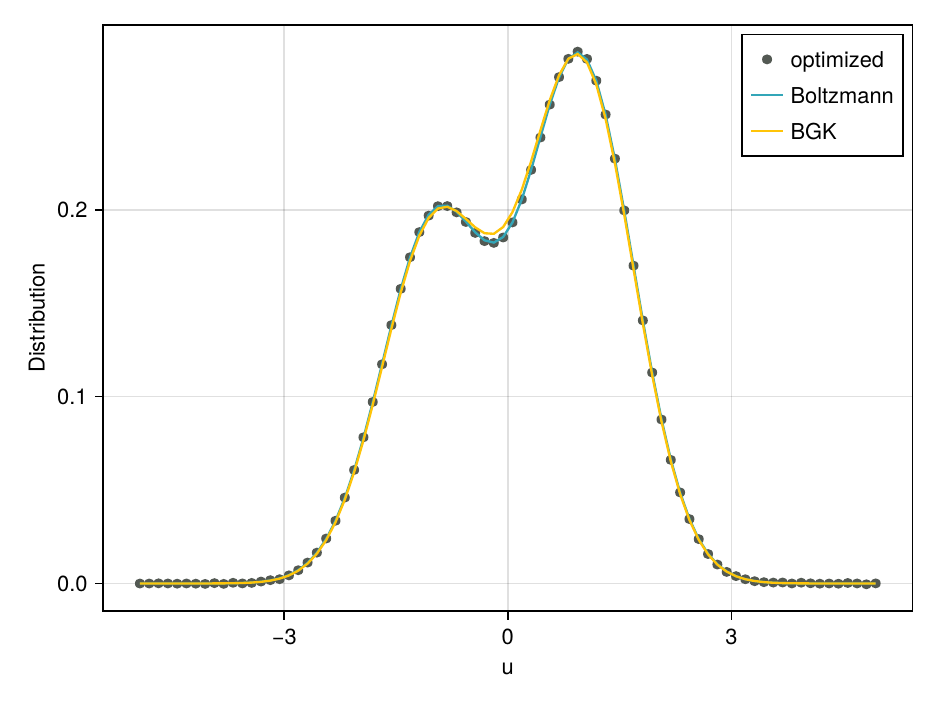}
    }
    \subfigure[$t=0.3$]{
        \includegraphics[width=0.47\textwidth]{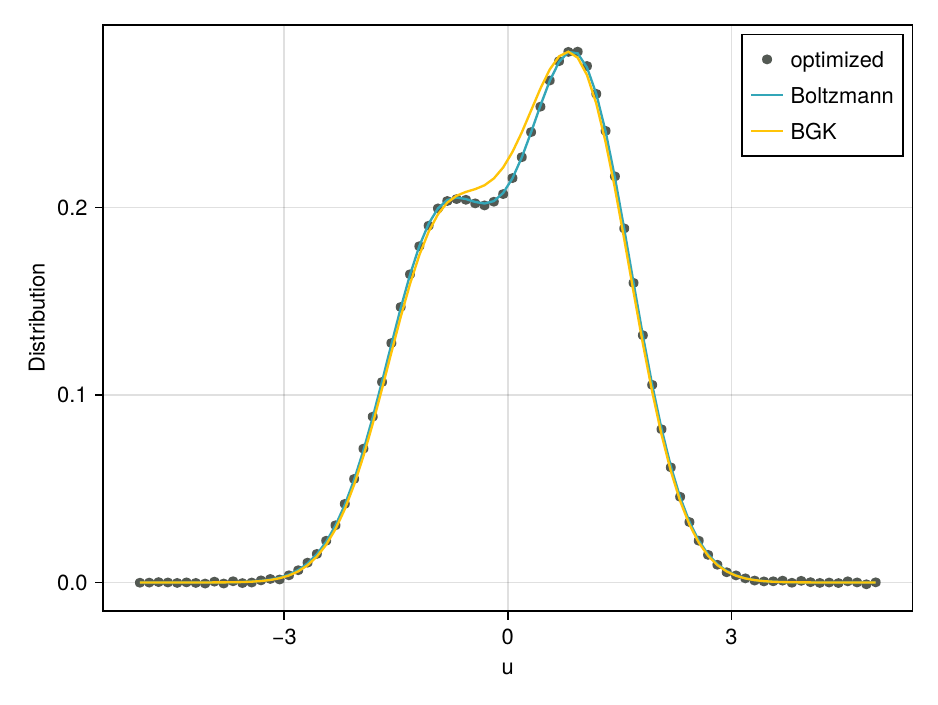}
    }
    \subfigure[$t=0.5$]{
        \includegraphics[width=0.47\textwidth]{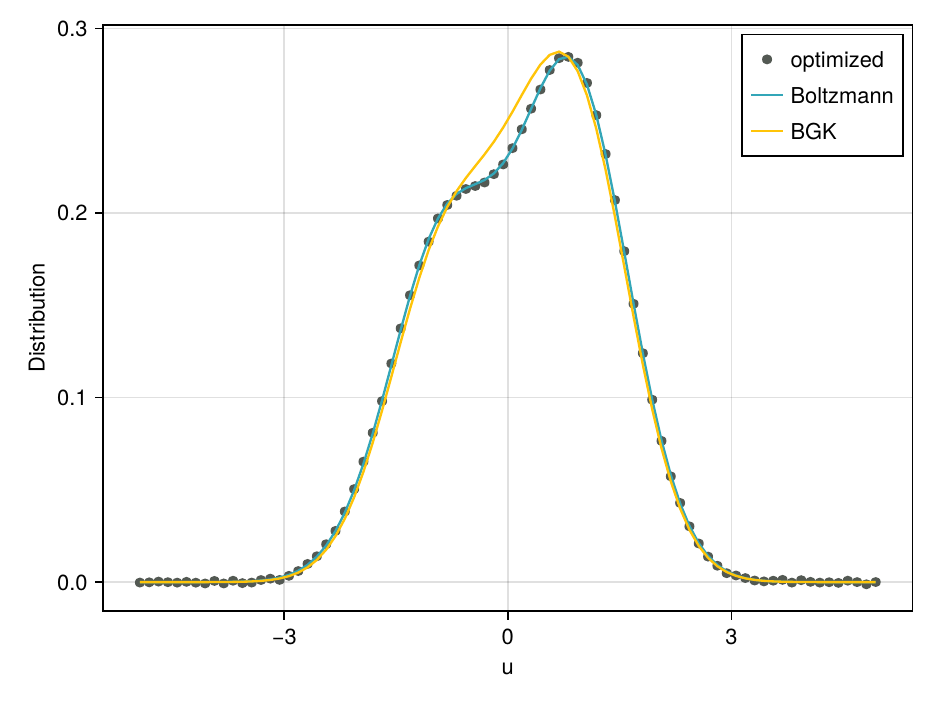}
    }
    \subfigure[$t=1.0$]{
        \includegraphics[width=0.47\textwidth]{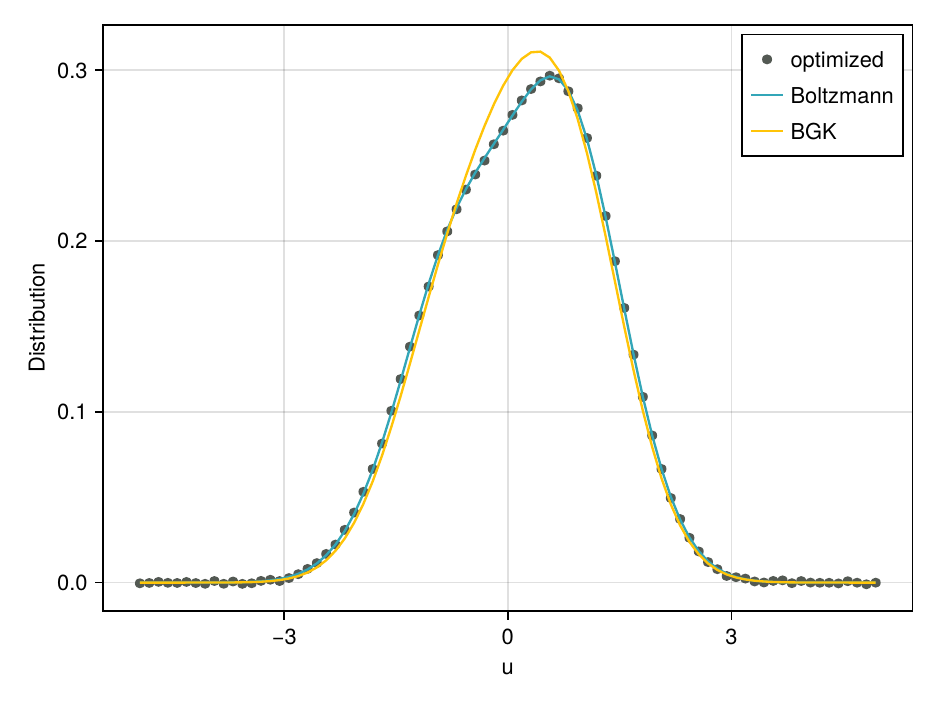}
    }
    \subfigure[$t=3.0$]{
        \includegraphics[width=0.47\textwidth]{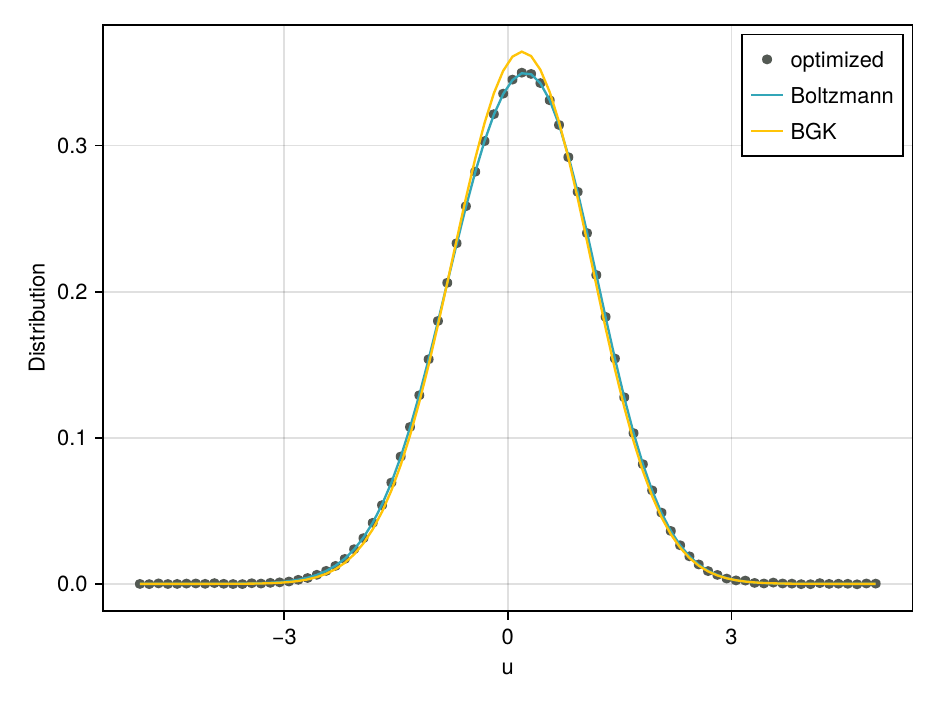}
    }
    \subfigure[$t=5.0$]{
        \includegraphics[width=0.47\textwidth]{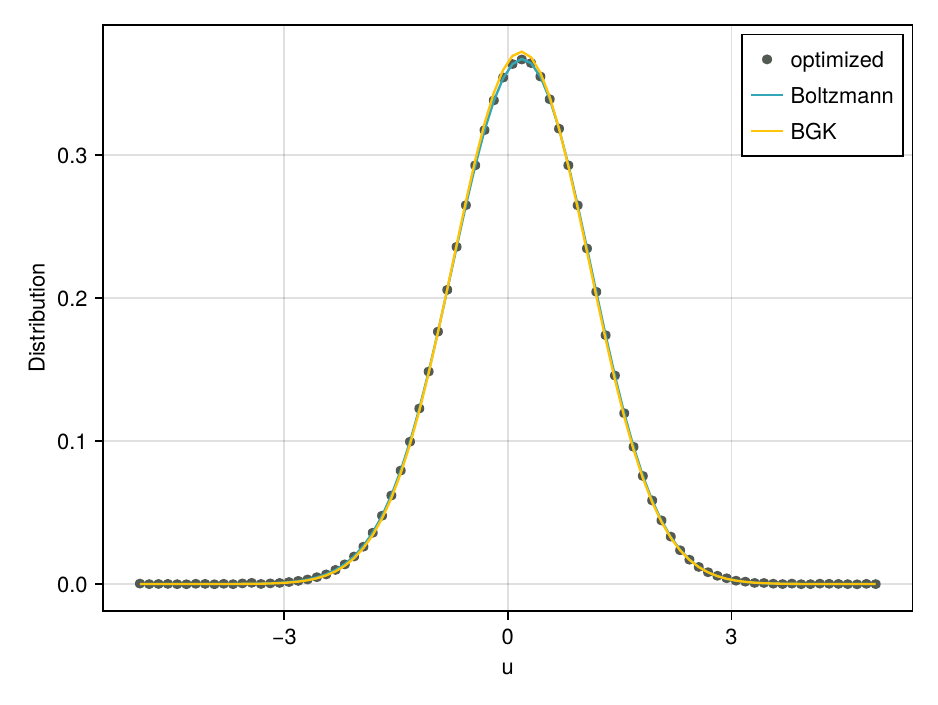}
    }
    \caption{Particle distribution functions at different time instants simulated by different models in the homogeneous relaxation problem (full Boltzmann equation as reference solution).}
    \label{fig:relax line}
\end{figure}

\begin{figure}[htb!]
    \centering
    \subfigure[$f$]{
        \includegraphics[width=0.47\textwidth]{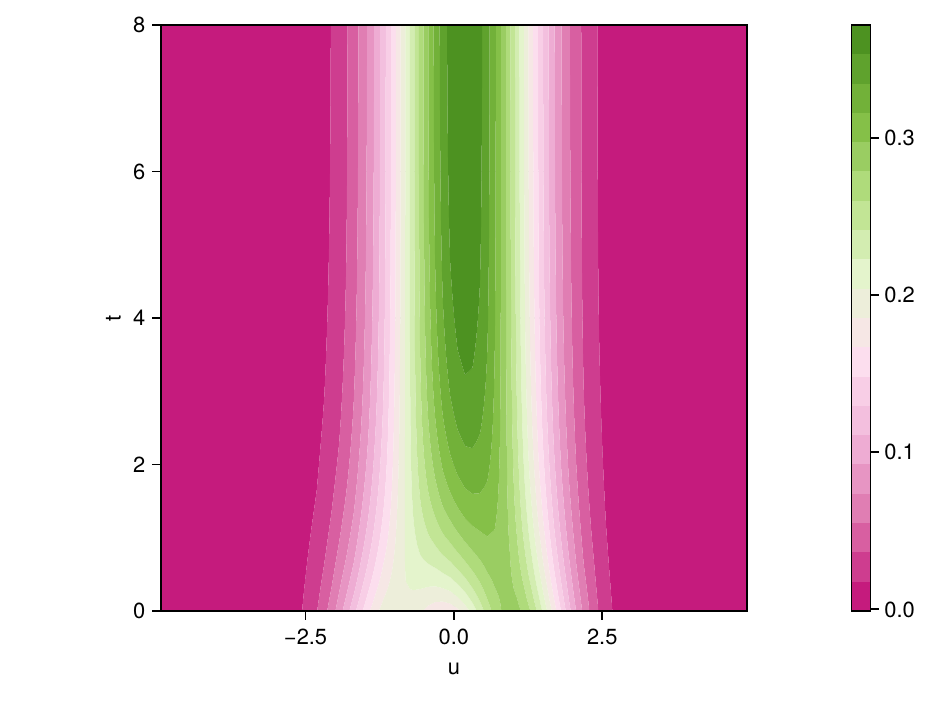}
    }
    \subfigure[$\mathcal Q(f)$]{
        \includegraphics[width=0.47\textwidth]{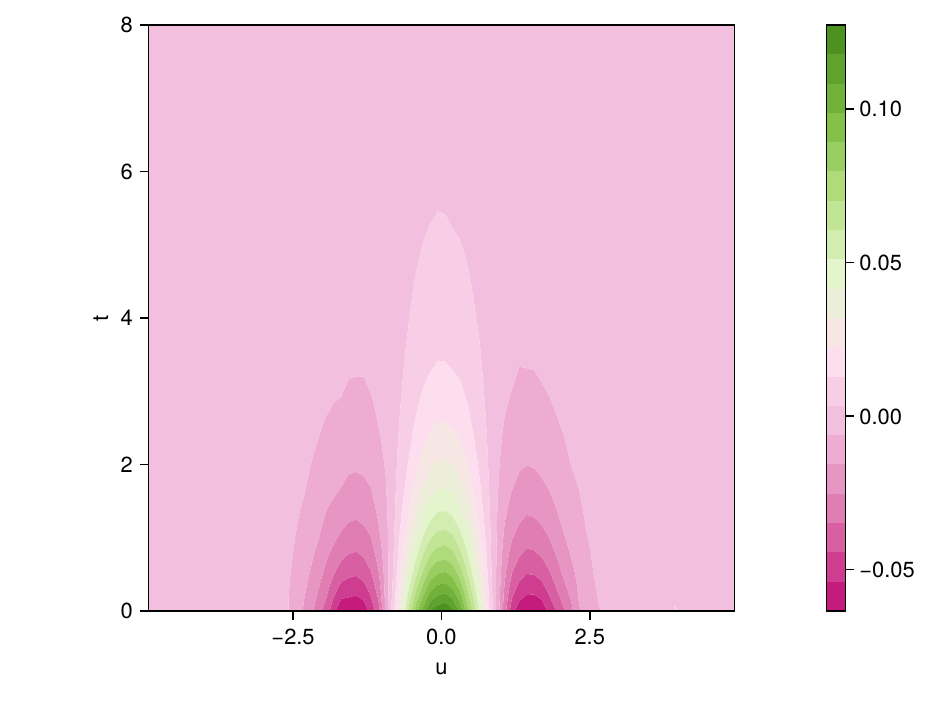}
    }
    \subfigure[$f-f_\mathrm{BGK}$]{
        \includegraphics[width=0.47\textwidth]{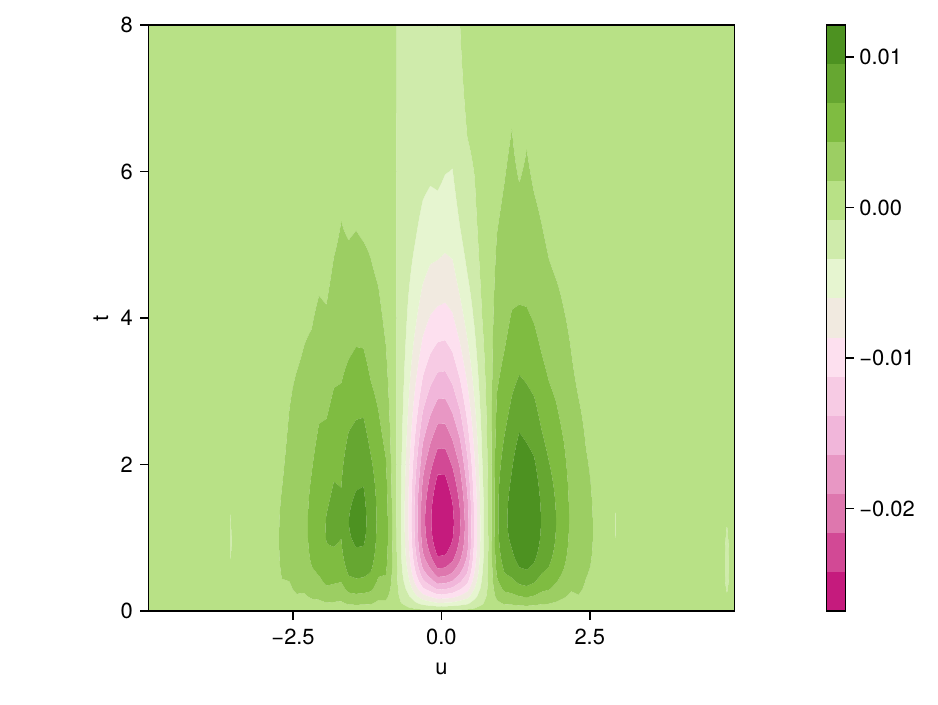}
    }
    \subfigure[$\mathcal Q(f) - \mathcal R(f)$]{
        \includegraphics[width=0.47\textwidth]{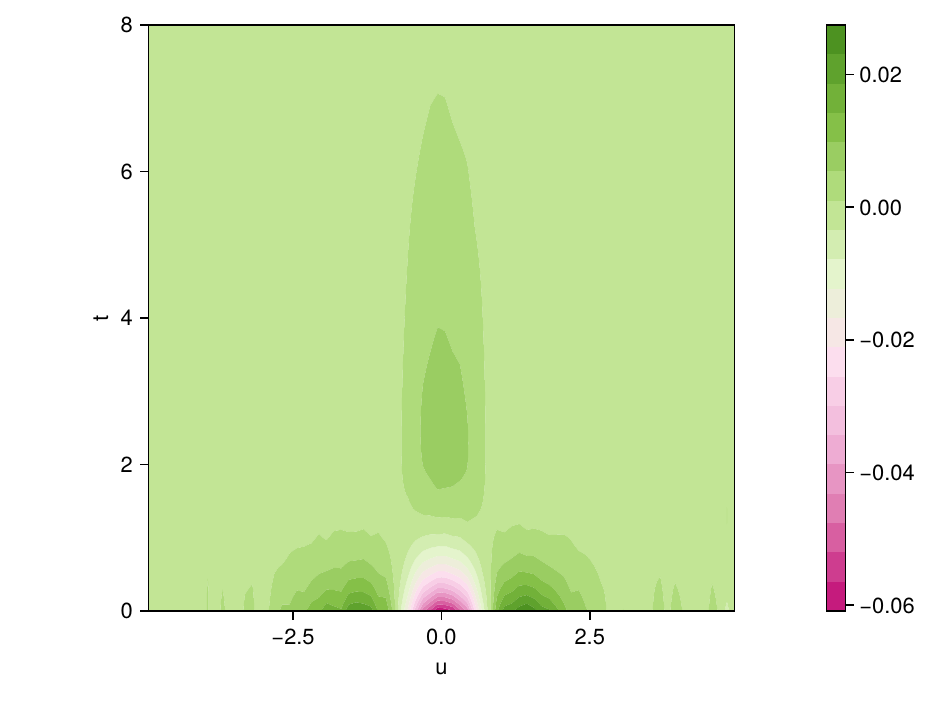}
    }
    \caption{Particle distribution functions, collision terms, and their differences over the time-velocity domain simulated by the DeepONet and BGK models in the homogeneous relaxation problem.}
    \label{fig:relax contour}
\end{figure}

Figure \ref{fig:relax line} presents the solutions of particle distribution function at different time instants simulated by the Boltzmann, BGK, and the current $\partial P$-based mechanical-neural model.
It can be seen that the current model outperforms the BGK model in terms of accuracy, and it provides a non-equilibrium evolutionary solution equivalent to the Boltzmann equation.
Figure \ref{fig:relax contour} provides the difference between the distribution function and the collision term provided by these two models over the entire time-velocity domain.
Clearly, it is the difference in the collision terms provided by the BGK equation (i.e., $\mathcal R(f)$) and the DeepONet neural model (i.e., $\mathcal Q(f)$) for the non-equilibrium distribution function that leads to the different solutions of the BGK and DeepONet models.
As the intermolecular interactions occur, the distribution function gradually approaches the equilibrium state, at which point the results of the two models converge.
Table \ref{tab:relaxation cost} presents the computational costs of a single simulation using the Boltzmann equation, the BGK model, and the $\partial P$-based mechanical-neural model.
Since the high-dimensional convolution operations in the fast spectral method are replaced by the tensor summations and products in the DeepONet, the computational cost is significantly reduced.
With the current parameter settings, the computational efficiency has improved by more than three orders of magnitude.
This numerical experiment verifies the ability of the $\partial P$-based solution algorithm to solve the Boltzmann equation accurately and efficiently.

\begin{table}[htbp]
    \caption{Computational setup of the relaxation problem.} 
    \centering
    \begin{tabular}{lllllllll} 
        \hline
        Equation & Gas & $t$ & $\mathbf v$ & $N_{v_x}$ & $N_{v_y}$ & $N_{v_z}$ \\ 
        Boltzmann & Argon & $(0,8]$ & $[-5,5]^3$ & 80 & 28 & 28 \\ 
        \hline
        Quadrature & Integrator & Kn & CFL & $N_p$ & Optimizer \\
        Rectangular & Tsit5 & 1 & 0.5 & $51600$ & LBFGS \\
        \hline
    \end{tabular} 
    \label{tab:relaxation}
\end{table}

\begin{table}[h]
    \centering
    \caption{Computational costs of a single simulation using the Boltzmann equation, the BGK equation, and the mechanical-neural model in the relaxation problem.}
    \begin{tabular}{lll}
        \toprule
        & Time ($10^{-3}$ s) & Allocation (MB) \\
        \midrule
        Boltzmann & $2.32\times 10^3$ & $6.35\times 10^3$ \\
        BGK & 0.90 & 1.37 \\
        $\partial P$ & 3.91 & 47.94 \\
        \bottomrule
    \end{tabular}
    \label{tab:relaxation cost}
\end{table}

\subsubsection{Normal shock wave structure}

We then turn to the normal shock wave structure problem.
Based on the reference frame of the shock wave, the initial flow field is set as
\begin{equation}
    \left(\begin{array}{c}
    \rho \\
    U \\
    T \\
    \end{array}\right)_{{t=0,x<0}}=\left(\begin{array}{c}
    \rho_- \\
    U_- \\
    T_- \\
    \end{array}\right), \quad 
    \left(\begin{array}{c}
    \rho \\
    U \\
    T \\
    \end{array}\right)_{{t=0,x\ge 0}}=\left(\begin{array}{c}
    \rho_+ \\
    U_+ \\
    T_+ \\
    \end{array}\right),
\end{equation}
where the subscripts $-$ and $+$ denote the upstream and downstream states of the shock wave, respectively.
The upstream and downstream conditions are related through the Rankine-Hugoniot relation, i.e.,
\begin{equation}
\begin{aligned}
    \frac{\rho_{+}}{\rho_{-}} &=\frac{(\gamma+1) \mathrm{Ma}^{2}}{(\gamma-1) \mathrm{Ma}^{2}+2}, \\
    \frac{U_{+}}{U_{-}} &=\frac{(\gamma-1) \mathrm{Ma}^{2}+2}{(\gamma+1) \mathrm{Ma}^{2}}, \\
    \frac{T_{+}}{T_{-}} &=\frac{\left((\gamma-1) \mathrm{Ma}^{2}+2\right)\left(2 \gamma \mathrm{Ma}^{2}-\gamma+1\right)}{(\gamma+1)^{2} \mathrm{Ma}^{2}},
\end{aligned}
\end{equation}
where $\rm Ma$ is the upstream Mach number, and the specific heat ratio takes $\gamma=5/3$ for a monatomic molecule.
The initial particle distribution function is set as Maxwellian in correspondence with the above macroscopic variables.
The reference state corresponds to the upstream conditions, and the system is non-dimensionalized by the upstream molecular mean free path.
The computational setup is detailed in Table \ref{tab:shock}.
The DeepONet model has a total of 20736 parameters. 
The cost function is defined as 
\begin{equation}
    C = \sum_i^{N_x}\sum_j^{N_v} |f_{i,j}^{\mathrm{ref}}(t=50)-f_{i,j}(t=50)|^2,
\end{equation}
where the reference solution $f^\mathrm{ref}$ is obtained by solving the Shakhov model equation with the same computational setup, which provides more accurate solutions for the evolution of high-temperature gases thanks to the heat flux-based correction.

\begin{figure}[htb!]
    \centering
    \subfigure[Density]{
        \includegraphics[width=0.47\textwidth]{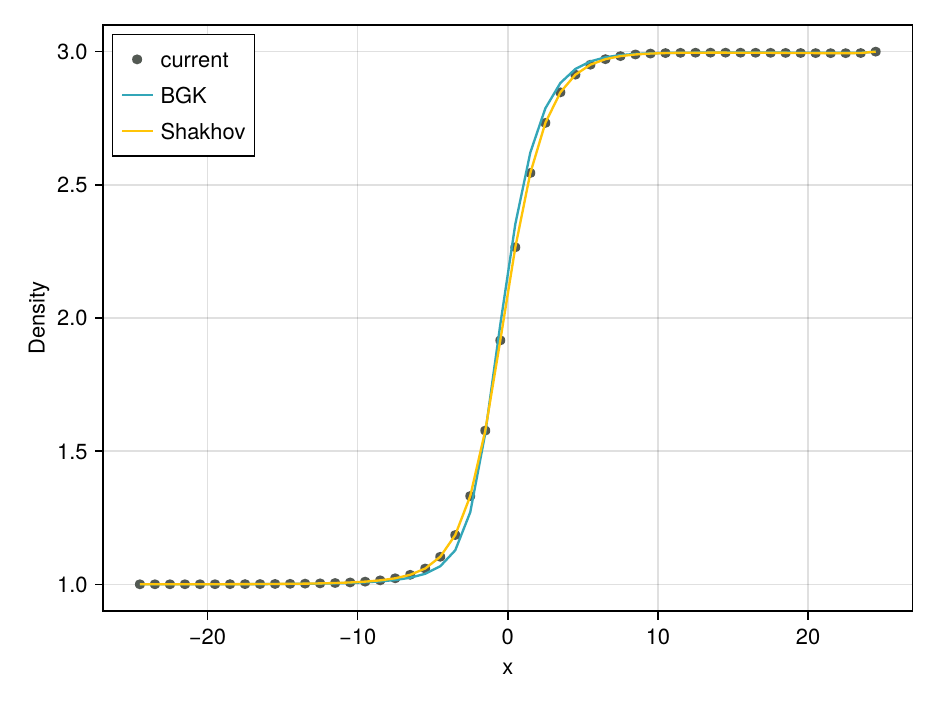}
    }
    \subfigure[Velocity]{
        \includegraphics[width=0.47\textwidth]{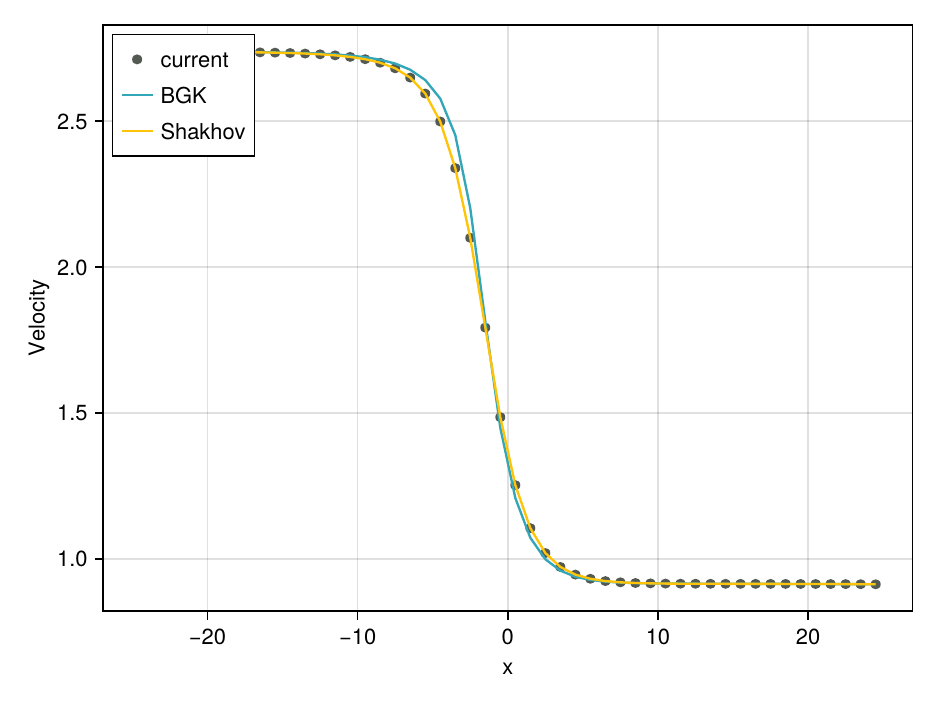}
    }
    \subfigure[Temperature]{
        \includegraphics[width=0.47\textwidth]{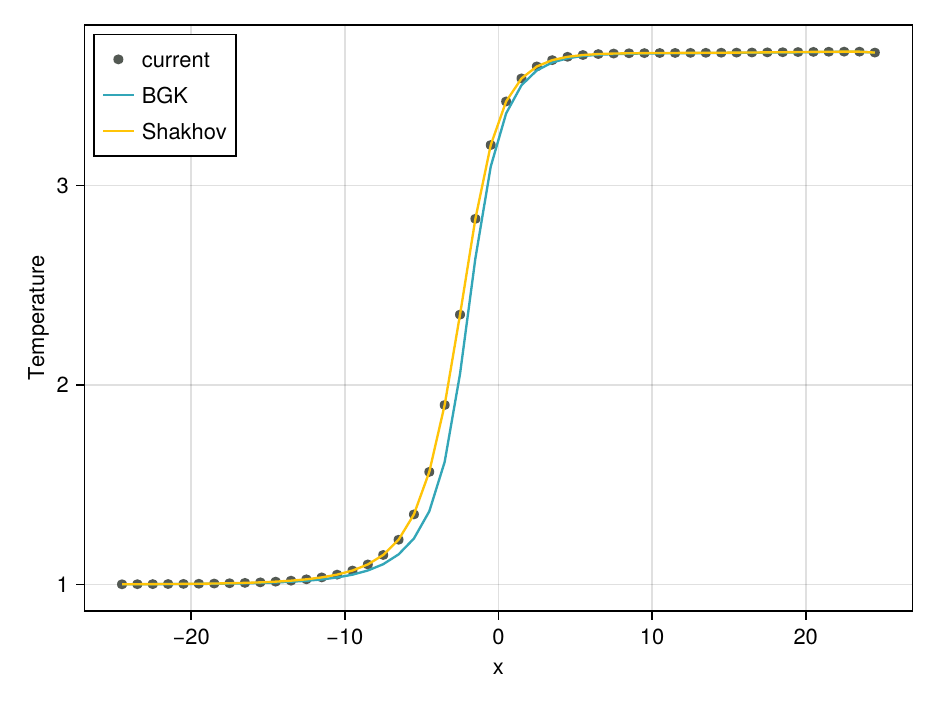}
    }
    \caption{Profiles of density, $U$-velocity, and temperature simulated by different models in the normal shock wave structure problem (Shakhov model as reference solution).}
    \label{fig:shock}
\end{figure}
\begin{figure}[htb!]
    \centering
    \subfigure[$f$]{
        \includegraphics[width=0.47\textwidth]{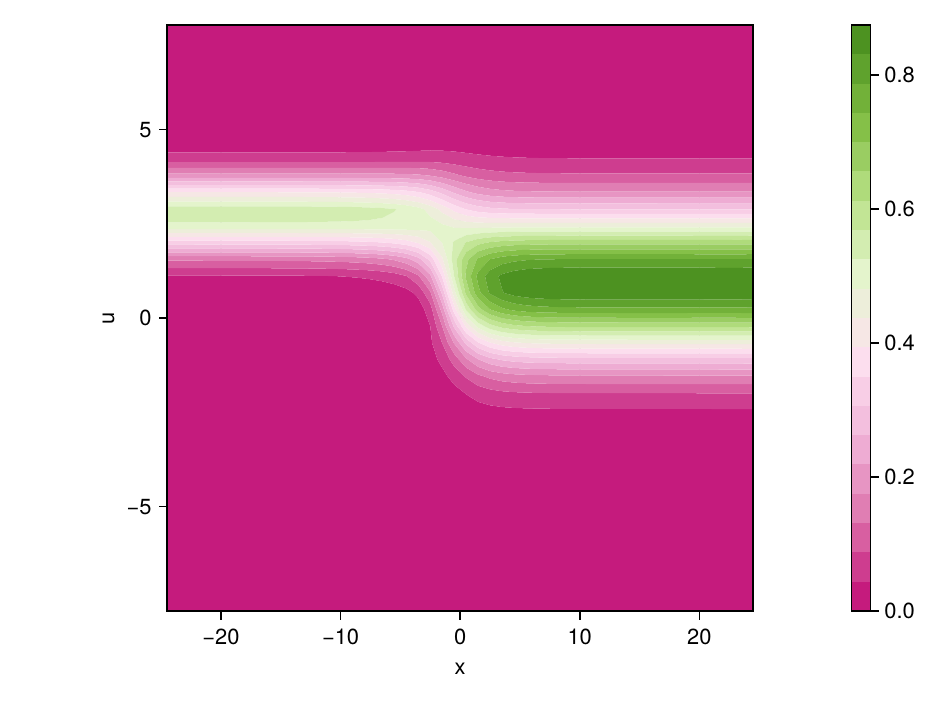}
    }
    \subfigure[$\mathcal Q(f)$]{
        \includegraphics[width=0.47\textwidth]{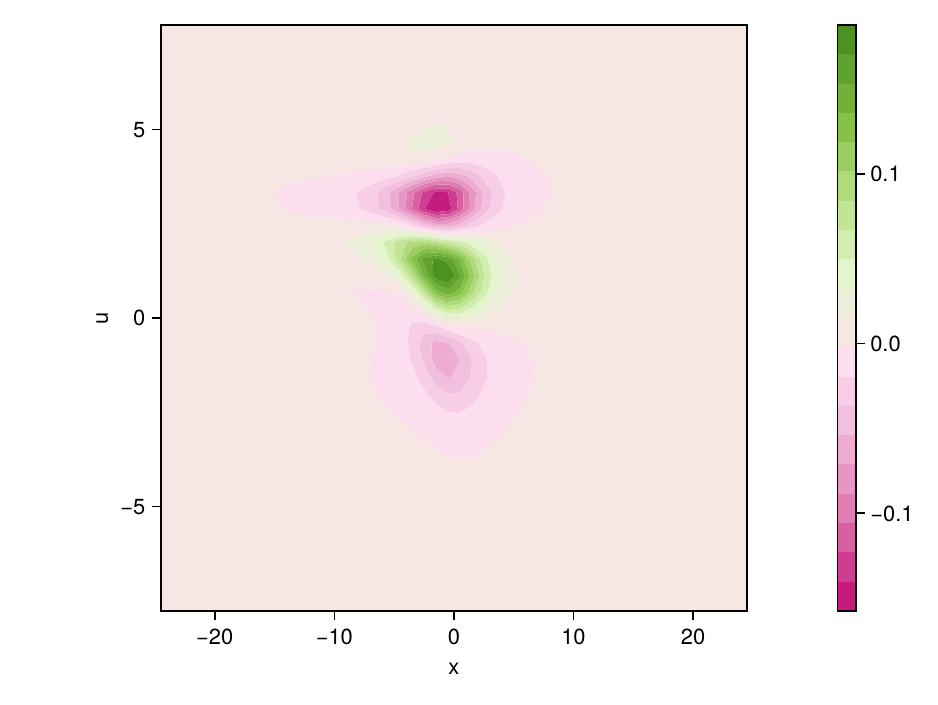}
    }
    \subfigure[$f-f_\mathrm{BGK}$]{
        \includegraphics[width=0.47\textwidth]{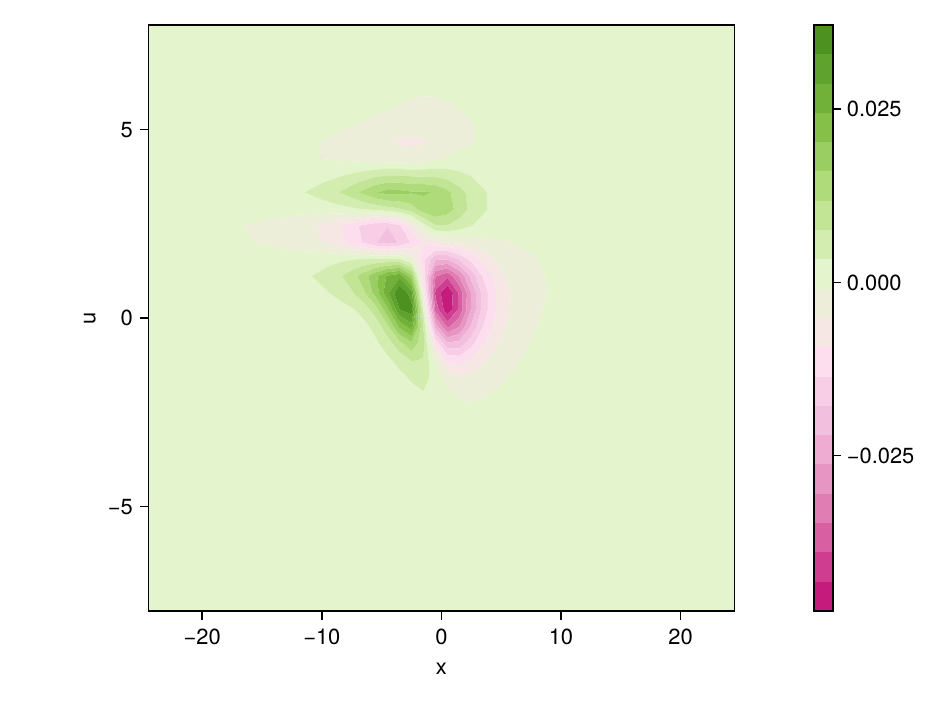}
    }
    \subfigure[$\mathcal Q(f) - \mathcal R(f)$]{
        \includegraphics[width=0.47\textwidth]{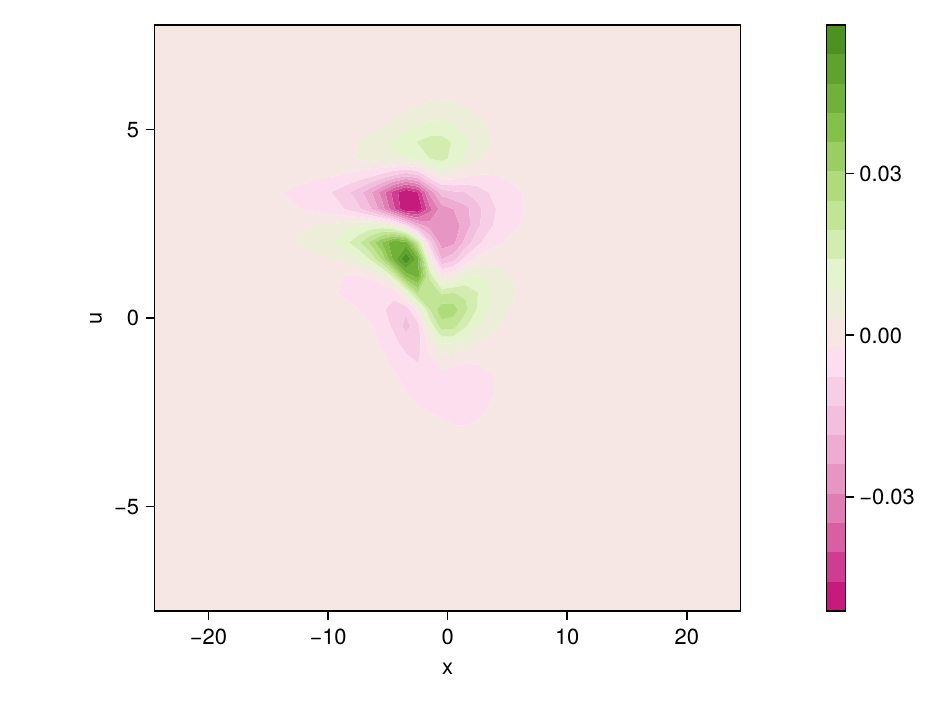}
    }
    \caption{Particle distribution functions, collision terms, and their differences over the space-velocity domain simulated by the unified mechanical-neural network model and BGK equation in the normal shock wave structure problem.}
    \label{fig:shock contour}
\end{figure}

Figure \ref{fig:shock} presents the distributions of density, velocity, and temperature simulated by the Boltzmann equation, the BGK equation, and the current mechanical-neural model.
The current model provides predictions that are equivalent to the reference solution.
Figure \ref{fig:shock contour} details the contours of distribution functions and collision terms for both models over the entire space-velocity domain.
As is shown, within the range of the shock wave (the width is of $O(10\ell)$), the flow variables change dramatically, leading to intensive intermolecular collisions.
Accurate prediction of collision effects is a prerequisite for capturing the correct shock profile.
The difference in the collision terms leads to different distribution functions and the corresponding macroscopic variables of the BGK and the mechanical-neural models.
This numerical experiment verifies the ability of the current $\partial P$-based solution algorithm to solve highly dissipative non-equilibrium flows with spatial inhomogeneity.

\begin{table}[htbp]
    \caption{Computational setup of the normal shock wave structure problem.} 
    \centering
    \begin{tabular}{lllllllll} 
        \hline
        Equation & Gas & $t$ & $x$ & $N_{x}$ & Order & $v$ & $N_{v}$ & Ma \\ 
        Boltzmann & Argon & $(0,50]$ & $[-25,25]$ & $50$ & 1 & $[-5,5]$ & 36 & 3 \\ 
        \hline
        Flux & Quadrature & Integrator & Boundary & CFL & $N_{p}$ & Optimizer \\
        Upwind & Rectangular & Euler & Dirichlet & 0.5 & $20736$ & AdamW \\
        \hline
    \end{tabular} 
    \label{tab:shock}
\end{table}

\section{Conclusion}

Research on multi-scale and non-equilibrium flows faces challenges arising from high dimensionality and strong nonlinearity in theoretical models and solution algorithms.
For the first time, this paper systematically addresses the application of differentiable programming to the construction of solution algorithms for multi-scale flow physics across continuum and rarefied regimes.
Leveraging composable automatic differentiation and adjoints, end-to-end optimization of key parameters in the models and algorithms can be seamlessly integrated with forward numerical simulation by computing gradients throughout the backward passes of the simulation program.
As a result, classical scientific computing and machine learning workflows are organically fused.
The paradigm of differentiable simulation lays a solid foundation for building unified mechanical-neural network models, enabling versatile data-driven approaches for physics discovery, surrogate modeling, and simulation acceleration.
It has great potential to be extended to the study of other complex systems, e.g., radiative transfer \cite{larsen2013properties,mcclarren2010robust}, plasma physics \cite{blaustein2024structure,von2014vlasiator}, stochastic simulation \cite{hu2017uncertainty,xiao2021stochastic}, and so on.



\section*{CRediT authorship contribution statement}

Tianbai Xiao: Conceptualization, Formal analysis, Investigation, Methodology, Project administration, Resources, Software, Visualization, Writing – original draft, Writing – review \& editing.

\section*{Declaration of competing interest}

The author declares that there are no known competing financial interests or personal relationships that could have appeared to influence the work reported in this paper.

\section*{Acknowledgement}

The current research is funded by the National Science Foundation of China (No. 12302381) and the Chinese Academy of Sciences Project for Young Scientists in Basic Research (YSBR-107).
The computing resources provided by Hefei Advanced Computing Center and ORISE Supercomputer are acknowledged.

\clearpage
\newpage

\bibliographystyle{unsrt}
\bibliography{main}
\newpage

\end{document}